\documentclass[a4paper,10pt]{article}
\pdfoutput=1
\usepackage[utf8]{inputenc}
\usepackage{amsmath}
\usepackage{graphicx}
\usepackage{caption}
\usepackage{subcaption}
\usepackage{amssymb}
\usepackage{color}
\usepackage{cancel}
\usepackage{framed}
\usepackage{comment}
\usepackage[margin=1in]{geometry}
\usepackage{mathtools}
\usepackage{hyperref}
\usepackage{chngcntr}

\usepackage[square,numbers,sort&compress]{natbib}

\counterwithin{equation}{section}

\newcommand{\ddx}[2]{\frac{\partial #1}{\partial #2}}

\newcommand{\nn}{\nonumber}

\def\a{\alpha}   
\def\d{\delta}
 \def\m{\mu}   \def\r{\rho}
 \def\o{\omega}  \def\l{\lambda}
\def\D{\Delta} \def\G{\Gamma}

 \newcommand{\Ocal}{{\mathcal O}}
 \newcommand{\Dcal}{{\mathcal D}}
 
 \newcommand{\Ical}{{\mathcal I}}

\newcommand*{\affaddr}[1]{#1} % No op here. Customize it for different styles.
\newcommand*{\affmark}[1][*]{\textsuperscript{#1}}

\def\Cincy{\small{Department of Physics, University of Cincinnati, Cincinnati, Ohio 45221, USA}}
\def\Weizmann{\small{Department of Particle Physics and Astrophysics, Weizmann Institute of Science, Rehovot 761001, Israel}}

\begin{document}

% Don't want date printed
\date{}
% Make title large and bold
\title{\Large\bfseries Stability of Condensed Fuzzy Dark Matter Halos}

% Target typesetting:
%
% Author A, Author B, Author C, Author D and Author E
%        A,B,C Department of Computer Science
%       D,E Department of Mechanical Engineering
%          Email A,B,C,D,E @university.edu
%                  Latex University

\author{%
Joshua Eby\affmark[$\dag$],
Madelyn Leembruggen\affmark[*],
Peter Suranyi\affmark[*], and
L.C.R. Wijewardhana\affmark[*]\\
{\it\affaddr{\affmark[$\dag$]\Weizmann}}\\
{\it\affaddr{\affmark[*]\Cincy}}\\
}

\begin{titlepage}
\maketitle
\thispagestyle{empty}

\vskip 2 cm

\begin{abstract}
Stability properties of gravitationally bound condensates composed of ultralight axionic Fuzzy Dark Matter (FDM) are studied.  Previous work has shown that astrophysical collisions could make self-gravitating condensates structurally unstable, making them prone to collapse and decay; in the context of FDM, we reexamine the relevant timescales using the time-dependent variational method. We show that FDM condensates can be made unstable through gravitational interactions with central black holes, for black hole masses in a phenomenologically relevant range. Instability could also be stimulated by galaxy collisions. The subsequent decay takes place over a period lasting as long as many thousands of years. We also discuss the possible relevance of FDM condensates to understanding the composition of Ultracompact Dwarf (UCD) Galaxies. Future observation of extremely massive black holes in the central regions of UCDs can constrain this interpretation.
\end{abstract}

\end{titlepage}

\maketitle

\section{Introduction}

Non-observation of weakly interacting massive particles (WIMPs), the most popular particle physics candidate for dark matter,  in laboratory experiments, has led many investigators to look for other viable candidates of dark matter. It has been suggested that cold dark matter models based on WIMPs work very well at large scales, but face difficulties describing data close to the cores of galaxies; see e.g. \cite{Weinberg}. There is some debate about the seriousness of these supposed discrepancies \cite{NoCuspCore,NoMissSat}, but regardless of the outcome of this debate, one may be sufficiently motivated at the present time to pursue alternative models. One avenue of research has been to investigate the viability of the hypothesis that macroscopic Bose-Einstein condensates (BECs) of spin zero bosons be the source of dark matter. Condensate dark matter may be less susceptible to the usual phenomenological difficulties of cold particle dark matter models at short distance scales, like the cusp-core or missing satellite problems.\footnote{This too has been called into question recently; see e.g. \cite{NoFDMCuspCore}.}

It has been established that spin zero particle excitations can form gravitationally bound dilute condensate bubbles, which are minima of the energy landscape \cite{Kaup,RB,BreitGuptaZaks,CSW,SS,Friedberg,SS2,Liddel,Lee}. Some early work in this field, emphasizing the cosmological evolution of scalar field perturbations, was performed by \cite{Khlopov1,Khlopov2,Khlopov3,Khlopov4,Khlopov5}. Recently, there has been a revival of this idea in the literature as it applies to well-motivated models of axion like particles \cite{Iwazaki,SikivieYang,ChavanisMR,ChavanisMR2,BB,BarrancoNS,TkachevFRB,ESVW,Guth,Braaten,ELSW,ELSW2,WilczekASt,ChavanisPT,Kling1,Kling2,Hertzberg1,sarkar,Hertzberg2,Hertzberg3}. These condensates could be formed from miniclusters in the early universe \cite{HoganRees,KolbTkachev}, as the typical relaxation time through gravitational interactions is shorter than the age of the universe \cite{TkachevSim}. If the energy scale of inflation is lower than $f$, the axion decay constant (or scale of symmetry breaking in the early universe), minicluster formation may be washed out,\footnote{Note that this conclusion may not hold if there is a period of early matter domination in the universe; see the recent works \cite{Nelson,Visinelli}.} in which case the dominant mechanism for axion star formation is likely to be direct collapse. In any case, we will not concern ourselves with the mechanism for formation in this work.

Axions typically have attractive self interactions at the leading order, and beyond a critical mass, axionic condensates become structurally unstable \cite{ChavanisMR,ChavanisMR2}. Recently, the present authors have investigated the collapse of condensates made of QCD axions when their mass becomes supercritical. A preliminary study was done using the standard low energy effective potentials for QCD axions: the instanton \cite{ELSW} and chiral \cite{ELSW2} potentials. In both cases, this nonrelativistic study indicated that such supercritical objects collapse to attain much smaller radii, but would be stopped by short-range repulsive interactions before reaching the Schwarzschild radius. Previous studies, including only the leading self-interactions, found similar collapse times but interpreted the endpoint as a black hole \cite{ChavanisCollapse}. 

By including relativistic effects, we further predicted \cite{ELSW} that a collapsing axion condensate would rapidly emit relativistic scalar particles as it collapsed, an effect known as a Bosenova in condensed matter literature \cite{Bosenova}. This effect is made possible by the rapidly increasing binding energy of the axions in the condensate during collapse, which gives rise to a very large classical decay rate \cite{ESW,EMSW}, and the prediction of a Bosenova was subsequently confirmed by numerical studies \cite{Levkov,Helfer,Michel}. This decay process (known as classical decay, because it proceeds through tree-level diagrams) is also well-known in the literature on \emph{oscillons}, a closely related structure formed from scalar particles \cite{Hertzberg,MTY,EMTWY}

QCD axions have masses typically in the range $m\sim 10^{-3}-10^{-6}$ eV, which implies decay constants of roughly $f\sim 10^{10}-10^{13}$ GeV, and the condensates they form have the typical mass and size of an asteroid. If these axion BEC bubbles, often termed axion stars, are a component of dark matter, then as they populate and roam around galaxies at viral velocities, they could collide with each other and also with other stars \cite{Cotner,ELLSW}. During such collisions, otherwise stable axion stars could become supercritical and unstable, leading to collapse and subsequently decay from number changing particle interactions. In a recent publication, the present authors have analyzed collapses of axion stars during such  collisions~\cite{ELLSW}. The collision rates presented in that work applied to QCD axion stars, though it more generally represented a novel mechanism for the destruction of otherwise stable condensates.

It is also interesting to study bubbles of axion like particles with much larger condensate sizes. A class of such models has recently received much attention \cite{Turner,Press,Sin,Hu,Goodman,Peebles,Amendola,Shapiro,Schive,Marsh,Witten,JLee}, which is an example of so-called Fuzzy Dark Matter (FDM). Axions of this type can form condensates with sizes in the $100$ parsec range, which are identified with the central cores of galaxies. In such a model, this axion BEC is said to be floating in a sea of virialized axion particles, based on numerical simulations which have been developed in the last several years \cite{Sim1,Sim2,Sim3,Sim4,Sim5}. In this paper, we apply the same calculational methods used in our previous works \cite{ELSW,ELLSW} to study condensates made of axions in the typical FDM mass range of $m\sim 10^{-20} - 10^{-22}$ eV. We will review the predictions of FDM theory and simulation, and analyze the interplay between the central FDM soliton and other astrophysical sources, such as black holes.

The idea of FDM has drawn quite a bit of attention recently, with some claiming that the favored mass range above is in tension with observations of the Lyman-$\a$ forest \cite{LymanAlpha} or the circular velocity of stars near the cores of galaxies \cite{BarBlum}. These analyses are based on simulations which will continue to improve, and could determine the fate of FDM as a viable model. In the meantime, it is interesting to continue to analyze the consequences of this class of models. 

Beginning in the next section, we will review the nonrelativistic formulation employed in our previous papers \cite{ELSW,ELSW2} to analyze low energy configurations of axions, and we will briefly acknowledge the role of relativistic effects during collapse. In Section \ref{sec:Numerical}, we will discuss the application to FDM halos; we will also analyze different physical systems in which such collapses might be stimulated. In Section \ref{sec:Decay}, we will show numerical results for, and analyze the subsequent decay of, a collapsing FDM condensate, comparing it with previous results found in the QCD case. We conclude in Section \ref{sec:Conclusions}. In the Appendix, we also analyze the instability of a metastable axion condensate due to tunnelling effects.

We will use natural units throughout, where $\hbar = c = 1$.

\section{Analytic Formulation} \label{sec:Analytic}

\subsection{Dilute and Dense States}
In this section we review the formalism used in our previous work \cite{ELSW}, which can be used to perform order of magnitude estimates for the mass and size of axion structures, and analyze their stability. To describe the axion self-interactions, we take the usual instanton potential for the axion field $\Phi$,
\begin{equation} \label{eq:Vaxion}
 V_a(\Phi)=m^2\,f^2 \left[1-\cos\left(\frac{\Phi}{f}\right)\right].
\end{equation}
It is typical to neglect the self-interactions of very light axions, especially in FDM, which is appropriate over a large range of parameters when $\Phi \ll f$ (see e.g. \cite{Witten}). However, in this work, self-interactions play an important role and cannot be neglected.

There are a number of motivations for including the effect of self-interactions. Even though the leading-order $\Phi^4$ coupling is $m^2/f^2 \sim 10^{-94} \lll 1$ (for typical FDM input parameters $m=10^{-22}$ eV and $f=10^{16}$ GeV), the number of particles in the condensate can be as large as $N \sim M_P\,f / m^2 \sim 10^{99}$ (where $M_P=1.22\times10^{19}$ GeV is the Planck mass), compensating the smallness of the coupling. 
We can make this point more quantitative by a closer examination of the core-halo relation \cite{Sim1,Sim2,BarBlum}
\begin{equation} \label{corehalo}
 M = 1.4\times10^{9} M_\odot \left(\frac{10^{-22} \text{ eV}}{m}\right)\left(\frac{M_{halo}}{10^{12}M_\odot}\right)^{1/3},
\end{equation}
which is an empirical relation coming from FDM simulations, which relates the halo mass $M_{halo}$ to the mass $M$ of the FDM soliton core. But given that the maximum mass of a self-interacting axion condensate is \cite{ChavanisMR,ESVW}
\begin{equation} \label{Mcritical}
 M_{c} \approx 10\,\frac{M_P\,f}{m} 
 		= 10^{10} M_\odot \left(\frac{f}{10^{16} \text{ GeV}}\right)\left(\frac{10^{-22} \text{ eV}}{m}\right),
\end{equation}
it is plausible that in very massive galaxies, the central soliton is not more than an order of magnitude from the maximum stable mass. In this regime the effect of self-interactions is not small. It has also been pointed out that even the extremely small self-interaction coupling of FDM is important in considerations of large-scale structure \cite{Riotto}, and affects the mass-radius relation for axion condensates as well \cite{EMSW}. One final note is that it is not possible to analyze dense configurations without taking into account large contributions from both attractive and repulsive terms in the potential (\ref{eq:Vaxion}) \cite{Braaten,ELSW}. For all of these reasons, we do not neglect self-interactions in this work.

In \cite{ELSW}, we calculated the total energy of an axion condensate in the nonrelativistic limit. We used a rescaling of the radius and particle number $R$ and $N$ in terms of dimensionless parameters $\rho$ and $n$,
\begin{equation} \label{eq:scaling}
 R = \frac{1}{m} \frac{\rho}{\sqrt{\delta}}, \quad \quad
 N = \frac{f^2}{m^2} \frac{n}{\sqrt{\delta}},
\end{equation}
where $\delta \equiv f^2 / M_P{}^2$, which gave a rescaled energy of
\begin{equation} \label{eq:energy}
 e(\rho) \equiv \frac{E(\rho)}{m\,N\,\delta} 
	=  \frac{D_2}{2C_2} \frac{1}{\rho^2} 
	    - \frac{B_4}{2 C_2{}^2} \frac{n}{\rho} - \frac{n}{\rho^3}v.
\end{equation}
The first two terms in eq. (\ref{eq:energy}) are the contributions of the kinetic and gravitational energy, respectively, while the third term gives rise to self-interaction terms,
\begin{equation} \label{eq:selfints}
 v = \sum_{k = 0}^\infty \left(-\frac{1}{2 C_2}\right)^{k+2} 
      \left(\frac{n\,\delta}{\rho^3}\right)^k 
	  \frac{C_{2k + 4}}{\left[(k+2)!\right]^2}.
\end{equation}
The coefficients $B_4$, $D_2$, and $C_k$, defined in \cite{ELSW}, are numerical constants that depend on the precise form of the condensate wavefunction (we give numerical values in Section \ref{sec:Gaus}). 

Assuming leading order of $\delta\ll1$ is an appropriate expansion, there exists a metastable minimum of the energy at a radius of
\begin{equation} \label{eq:rhodilute}
 \rho_d = \frac{C_2\,D_2}{B_4\,n}\left(1 - \sqrt{1 - \frac{n^2}{n_c{}^2}}\right)
\end{equation}
with a dimensionless critical particle number of
\begin{align} \label{eq:nCrit}
 n_c &= \sqrt{\frac{8}{3}} \frac{C_2 D_2}{\sqrt{B_4 C_4}} 
\end{align}
and corresponding radius
\begin{equation} \label{eq:rhoCrit}
 \rho_c = \sqrt{\frac{3 C_4}{8B_4}}
\end{equation}
For particle number $N > N_c$, the energy functional no longer possesses this local energy minimum at $\rho = \Ocal(1)$, a configuration known as a \emph{dilute axion star}. Thus the dilute axion star has a maximum mass of 
\begin{equation} \label{eq:Mmax}
 M_c = m\,N_c = \frac{M_P\,f}{m}\,n_c.
\end{equation}

\begin{figure}
 \centering
 \includegraphics[scale=.85]{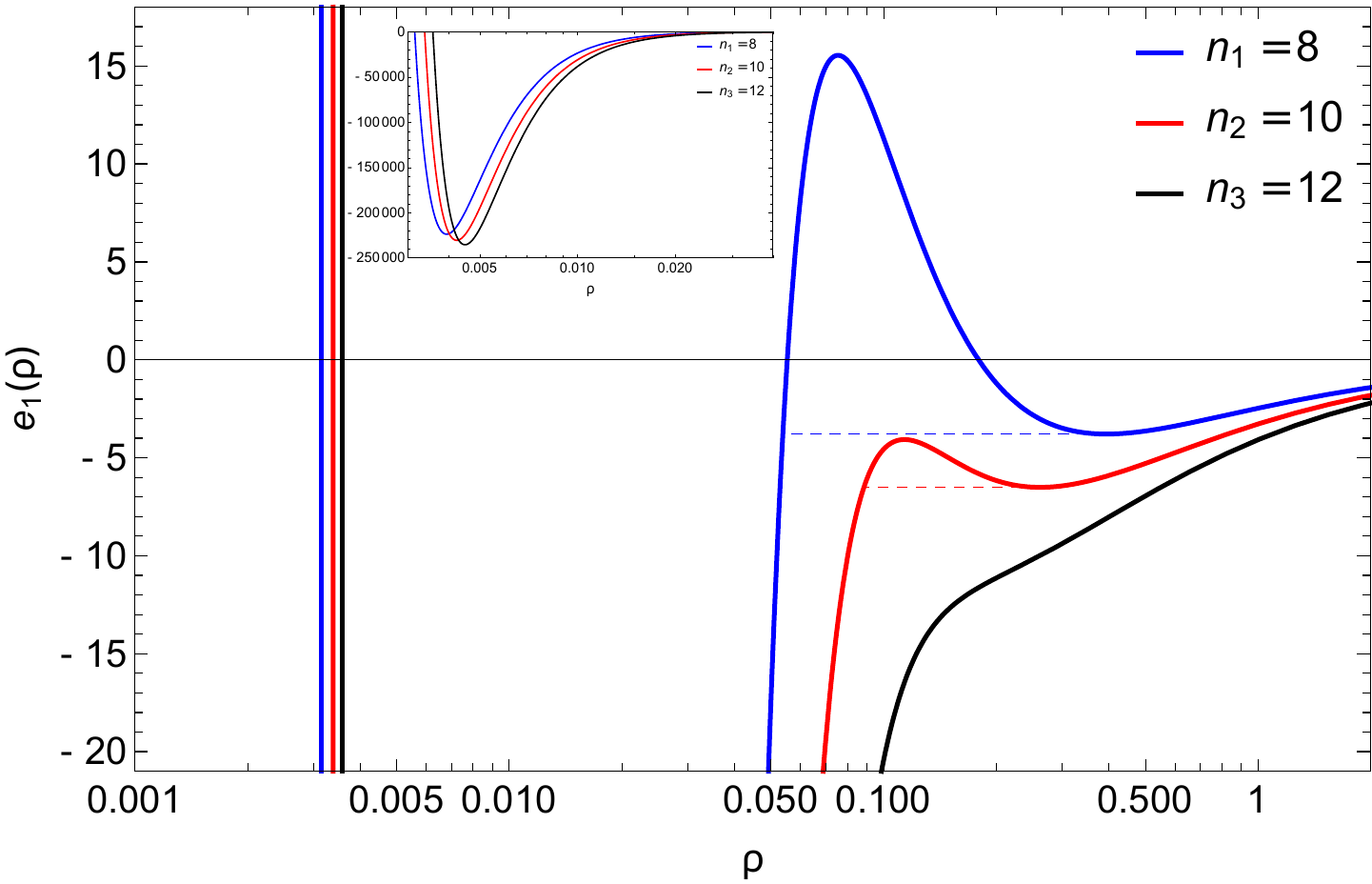}
 \caption{The rescaled energy of an axion star, eq. (\ref{eq:e1}), as a function of the rescaled radius $\r$, for three choices of rescaled particle number $n=8,10,12$, calculated using the Gaussian ansatz introduced in Section \ref{sec:Gaus}. The dilute minimum is represented in the main portion of the figure, near $\r=\Ocal(0.1-1)$, and the dense minimum can be seen in the inset at a much smaller value of $\r$. The third (black curve) has a particle number above the critical value $n_c\approx10.88$, and so does not have a local minimum of the energy. The dashed blue and red lines connect the dilute minima $\r_d$ to their equal-energy counterpoint $\r_1$ (see Appendix \ref{AppA}).}
 \label{fig:Energy}
\end{figure}

However, the global minimum of the energy exists at $\rho\ll1$ \cite{ELSW}, which corresponds to a configuration called a \emph{dense axion star} \cite{Braaten}. In this regime, the self-interactions are actually dominant over both the gravitational and kinetic energy terms. A truncation at $k=1$ of eq. (\ref{eq:selfints}), which includes the leading attractive and repulsive terms, implies the following estimate for the radius of this global minimum:
\begin{equation} \label{eq:rhoD}
 \rho_D \approx \sqrt{\frac{2}{\pi}} \left(\frac{n \,\delta}{3^{7/2}}\right)^{1/3}.
\end{equation}
Both the dilute and dense axion star configurations are represented in Figure \ref{fig:Energy}, for three choices of particle number $N$. Note that on the dense branch, the nonrelativistic analysis breaks down \cite{Backreaction,GRB}, but the vast majority of the collapse process occurs in the nonrelativistic region.

There are a number of ways that a dilute axion star might collapse. The simplest scenario is that it could accrete additional particles until its mass has grown larger than $M_c$. In a previous work \cite{ELLSW}, we also showed that during astrophysical collisions, either the effective mass of the axion star increases (due to the overlap of two axions stars), or the effective critical mass decreases (through interactions with some other astrophysical body, like an ordinary star). In either case, an otherwise stable dilute axion star might collapse, and its radius would fall from $\rho_d$ towards $\rho_D$, the latter being the endpoint if we neglect relativistic effects. Along the way, the axion star decays and loses a significant fraction of its mass to relativistic axion emission \cite{ELSW,Levkov,Helfer,ChavanisPT,Michel}. It is this collapse from dilute toward dense configurations that we are concerned with in this work. This process has been analyzed previously for QCD axions \cite{ChavanisCollapse,ELSW,ELSW2}; in what follows we will discuss the physical collapse scenarios for FDM condensates, and how the results differ from the QCD case.

Before moving on, it is also interesting to note that since the dilute axion star configuration is a local minimum of the energy, it is susceptible in principle to tunnelling across the energy barrier towards the dense global minimum of the energy. This tunnelling process can be analyzed using a WKB formalism, taking into account instantons which connect these classically disconnected configurations. That this rate be very small is a precondition for the existence of an axion star, but to our knowledge it has not been computed for a gravitating axion condensate in the literature previously. We perform the relevant calculation in Appendix \ref{AppA} and show that dilute axion stars are indeed very stable against tunneling processes, the rate being strongly suppressed by the large number of particles $N$ in the condensate.

\subsection{Effect of Relativistic Corrections}

Dilute axion stars are well described using a nonrelativistic theory coupled to Newtonian gravity. On the so-called transition or dense branches of solutions, gravity is effectively decoupled from the theory, as self-interactions begin to dominate in this regime.\footnote{This is most easily seen by inspecting the energy functional in eq. (\ref{eq:energy}): at small values of $\rho$, the gravitational term $\propto \r^{-1}$ decouples first, before the kinetic energy $\propto\r^{-2}$ or the self-interaction $\propto\r^{-3}$. Thus the Thomas-Fermi approximation \cite{Braaten}, which neglects the kinetic energy but not the gravitational interaction, is not appropriate in this region.} Importantly, dilute axion stars do not collapse to form black holes unless $f/M_P \sim \Ocal(1)$ (for that case, see \cite{Helfer,ChavanisPT}). In the nonrelativistic analysis, this can be understood by the fact that on the dense branch, the condensate radius $R_D$ is larger than the Schwarzschild radius for nearly all values $n$ as long as $\d \lesssim 1$ \cite{ELSW,ELSW2}. Thus in the vast majority of parameter space, contributions from general relativity are completely negligible.

There are, however, important corrections coming from special relativity for dense states (which have large binding energy), and this is not taken into account in the formalism outlined above; this has been noted in various works, including \cite{Levkov,Hertzberg1,Backreaction}. As we will see below, the nonrelativistic limit is appropriate for most of the collapse process, though decay and other relativistic effects become important near the end (we also noted this fact in \cite{ELSW}). These corrections are nontrivial to calculate, and as a result, there are still many open questions regarding dense axion stars.

For example, it is not currently known whether dense states exist up to arbitrarily high masses. The original work on the topic \cite{Braaten} suggested that they do, but this work (as above) was based on fully nonrelativistic calculations, as described in \cite{WilczekASt}. In a recent paper, the authors of \cite{Hertzberg1} pointed out that the correct endpoint of field configurations should coincide with $\Phi \sim 2\,\pi\,f$, where the range of the relativistic potential of eq. (\ref{eq:Vaxion}) ends. This led these authors to the conclusion that dense states only exist over less than an order of magnitude in mass, cutting off at a maximum of less than $M\sim 1000 f^2/m$. While the endpoint in $\Phi$ is surely correct, the mass endpoint suggested is only correct if one neglects higher harmonic dependence of the wavefunction on the eigenenergy $\mu_0$ \cite{WilczekASt}. Indeed, at leading order there are corrections from virtual lines with energy $3\mu_0$. Taking these corrections into account, it was shown in \cite{GRB} by direct calculation of the distribution inside a dense axion star that these states exist at least up to nearly $M\sim 10^{6} f^2/m$; at that point the calculation becomes very computationally taxing, as corrections from modes with energy $5\mu_0$ and higher become important. This calculation was sufficient to confirm that the nonrelativistic relation $M \sim R^3$ holds over a few orders of magnitude on the dense branch, which was not seen in \cite{Hertzberg1}. The exact cutoff of masses on the dense branch remains an open question. In any case, for the purposes of this work, we will ignore these corrections, and we postpone a discussion of the relativistic decay processes to Section \ref{sec:Decay}.

\subsection{Gaussian Ansatz} \label{sec:Gaus}

As an example, we may assume the wavefunction is a Gaussian with the typical form 
\begin{equation}
 \psi (r) = \frac{\sqrt{N}}{\pi^{3/4}R^{3/2}} e^{-r^2/2\,R^2},
\end{equation}
in which case the coefficients in the expressions above are \cite{ELSW}
\begin{equation} \label{GausParams}
 D_2 = \frac{3 \pi^{3/2}}{2}, \quad B_4 = \sqrt{2 \pi^5}, \quad C_k = 2\sqrt{\frac{2 \pi^3}{k^3}}.
\end{equation}
Then the scaled energy functional is 
\begin{equation}
 e(\rho) = \frac{3}{4} \frac{1}{\rho^2} 
	  - \frac{1}{\sqrt{2\pi}} \frac{n}{\rho} 
	  - \frac{1}{2\delta} \sum_{k = 0}^\infty 
		  \frac{(-1)^k}{\left[(k+2)!\right]^2 (k + 2)^{3/2}} 
		  \left(\frac{n\,\delta}{2 \pi^{3/2} \rho^3}\right)^{k+1}.
\end{equation}
Truncating the self-interaction terms at $k =1$, we find the following expression:
\begin{equation} \label{eq:e1}
 e(\rho)\Big{|}_{k_{max}=1} \equiv e_1(\rho) = \frac{3}{4} \frac{1}{\rho^2} 
    - \frac{1}{\sqrt{2\pi}} \frac{n}{\rho} - \frac{1}{32 \pi \sqrt{2 \pi}} \frac{n}{\rho^3} + \frac{\delta}{864 \pi^3 \sqrt{3}} \frac{n^2}{\rho^6}.
\end{equation}
Using this ansatz we calculate the critical particle number and radius of eqs. (\ref{eq:nCrit}) and (\ref{eq:rhoCrit}) to be 
\begin{equation}
 n_c = 2\sqrt{3}\pi \approx 10.88, \quad \quad 
 \rho_c = \sqrt{\frac{3}{32 \pi}} \approx 0.173.
\end{equation}
Importantly, the physical radius $R_{99}$, inside which $0.99$ of the mass is contained, is related to the variational parameter approximately by $R_{99} \approx 3\,R$; this relationship depends on the choice of ansatz.

Ignoring any relativistic effects (including decay), we can calculate the total collapse time for a dilute axion star approaching the dense configuration. The classical collapse time from an initial radius $R_0$ to final radius $R_D$ is \cite{ChavanisCollapse,ELSW}
 \begin{align} \label{eq:CollapseTime}
  t_{collapse} &= \sqrt{\frac{\alpha\,M}{2}} 
	      \int_{R_D}^{R_0}\frac{dR}{\sqrt{E(R_0) - E(R)}} \nn \\
      &=\frac{M_P{}^2}{m\,f^2}\sqrt{\frac{\alpha}{2}}
	      \int_{\rho_D}^{\rho_0}\frac{d\rho}{\sqrt{e(\rho_0) - e(\rho)}},
 \end{align}
 where $\a = 3/4$ for the Gaussian ansatz \cite{ChavanisMR}. Our original application of eq. (\ref{eq:CollapseTime}) in \cite{ELSW} was a calculation of the collapse time for QCD axion condensates. Two things change when considering the FDM case. First, of course, the input parameters $m$ and $f$ change, resulting in an overall rescaling of eq. (\ref{eq:CollapseTime}). But secondly, the endpoint of the collapse $\rho_D$ is at a much larger value, as $\d$ is much larger in FDM compared to QCD (c.f. eq. (\ref{eq:rhoD})). Whereas the former change increases the collapse time, the latter actually tends to decrease it. It is necessary thus to go through the calculation again carefully; we will describe these results in Section \ref{sec:Decay}.

\section{Collapse Scenarios} \label{sec:Numerical}

\subsection{Black Holes}

%There are a number of scenarios in which collapsing FDM condensates are relevant. We discuss a few of these cases below.

Black holes generally play an important role in studies of axion stars.
For example, it has been suggested that supermassive black holes are formed through collapse of condensates composed of ultralight axion-like particles, like the ones in models of FDM we have analyzed here \cite{Gupta}. In the noninteracting case, it is well-known that there is a maximum mass $M_c^{NI} = 0.633 M_P{}^2/m$ above which no stable condensed state exists \cite{Kaup,RB}. Above this mass, a previously stable condensate will form a black hole. 

The picture is very different when self-interactions are included. In the case of the axion potential of eq. (\ref{eq:Vaxion}), the maximum mass of eq. (\ref{eq:Mmax}) is smaller than $M_c^{NI}$ by a factor of $f/M_P \ll 1$. However, in this case the collapse of the condensate is stopped prior to black hole formation by short-range repulsive forces in the potential \cite{Braaten,ELSW,ELSW2}. This is equivalent to the statement that $\rho_D$ is much greater than the corresponding Schwarzschild radius.

It is also instructive to point out the case of a pure attractive $\l\,\phi^4$ theory (i.e. $\l<0$). In this case, there appears to be no short-range force to stabilize the collapse, and one might be led to the conclusion that a black hole is formed at the end of the collapse \cite{ChavanisCollapse}. However, this result may not be correct either. Even if at leading order the only self-interaction is attractive, relativistic corrections generate effective self-interactions of both attractive and repulsive types. The leading relativistic correction to $\l\,\phi^4$ theory generates a term of the form $\l^2\,\phi^6$, which is repulsive \cite{MTY,GRB,GuthRelativistic,BraatenRelativistic,EMTWY}, and can possibly stabilize the potential. An analysis of this case, including relativistic corrections, is beyond the scope of this work.

Black holes are relevant to FDM systems for another reason. Many known galaxies, including the Milky Way and Andromeda, contain a supermassive black hole at their center, in addition to dark matter and baryonic matter. The additional gravitational effect of a black hole on the stability of a FDM soliton has not previously been considered, as simulations of FDM do not typically include this contribution (nor do they include the self-interactions that make collapse relevant). We can analyze this effect by computing the effective contribution to the energy from the black hole, which is
\begin{equation}
 e_{BHg} = \frac{E_{BHg}}{m\,N\,\d} = -\frac{2}{\sqrt{\pi}} \frac{\mu_{BH}}{\r},
\end{equation}
where $\mu_{BH} = M_{BH} m / f\,M_P$ is a rescaled black hole mass defined in analogy to eq. (\ref{eq:scaling}). We have assumed that the radius of the black hole is much smaller than the condensate radius $\rho_d$, which is certainly appropriate for applications to realistic galaxies. The total gravitational energy (self + black hole) for the condensate in the Gaussian ansatz is
\begin{equation}
 e_{g,tot} = - \left[ 1 + 2\sqrt{2}\frac{\mu_{BH}}{n} \right] \frac{1}{\sqrt{2\pi}} \frac{n}{\rho}.
\end{equation}
The effect is to shift the parameter $B_4$ in eq. (\ref{GausParams}) by the factor $(1 + 2\sqrt{2}\mu_{BH}/n)$, so that the critical particle number of eq. (\ref{eq:nCrit}) shifts as $n_c \propto (\mu_{BH} / n)^{-1/2}$. This implies
\begin{equation}
 n_c^{eff} = \frac{2\sqrt{3}\pi}{\sqrt{1 + 2\sqrt{2}\frac{\mu_{BH}}{n}}}.
\end{equation}
The solution of $n=n_{c}^{eff}$ is
\begin{equation}
 n_c^{eff} = \sqrt{2\mu_{BH}^2 + 12\pi^2} - \sqrt{2}\mu_{BH}.
\end{equation} 
This is to say, otherwise stable condensates are made unstable by sufficiently massive black holes. In particular, instability is stimulated if $N>N_c^{eff}$.

\begin{figure}[t]
\centering
 \includegraphics[scale=.75]{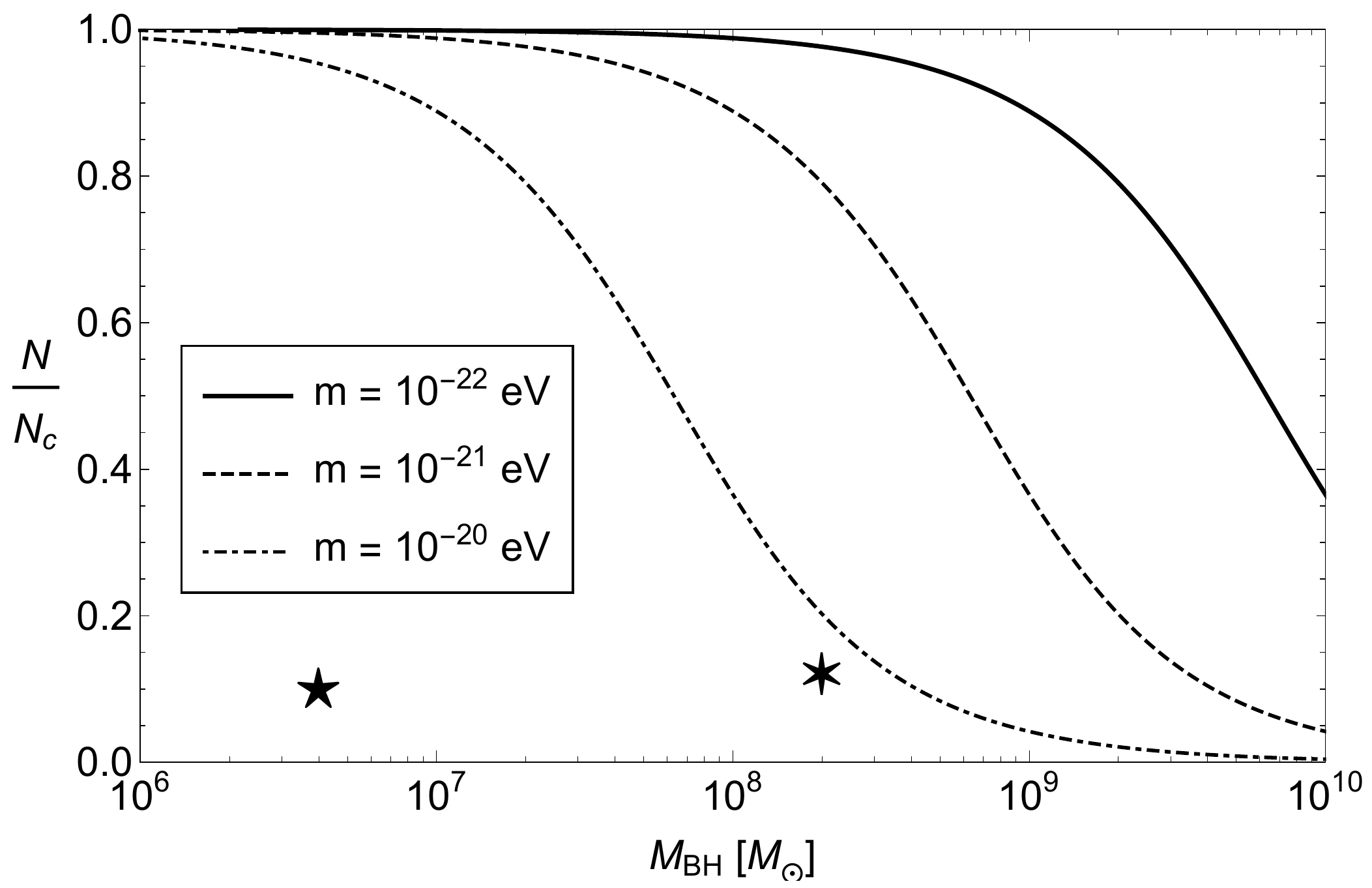}
 \caption{The fraction $N/N_c$ (vertical axis) vs. the mass of the black hole in the center of the condensate (horizontal axis). The curves represent the effective critical number $N=N_c^{eff}$ for a model with decay constant $f=10^{16}$ GeV and particle mass $m$ given in the legend. Regions to the right of these curves are made unstable to collapse by the black hole.  The critical particle number $N_c$ can be computed using $N_c = M_c / m$ and eq. (\ref{Mcritical}). The 5-pointed (6-pointed) star represents the expected soliton and black hole mass for the Milky Way (Andromeda) galaxy.}
 \label{fig:BHplot}
\end{figure}

In Figure \ref{fig:BHplot}, the curves represent the boundary of stability for $N=N_c^{eff}$ and the black hole mass $M_{BH}$, for different choices of the axion mass $m$. A condensate of particle number $N$ enshrouding a black hole of mass $M_{BH}$ will be stable if it lives to the left of the corresponding curve; a configuration to the right of the curve will collapse. The right edge of the plot corresponds approximately to a black hole of mass $10^{10}M_\odot$, which is in the range of the most massive black holes known to date (see e.g. \cite{HeavySMB}).

This scenario can potentially provide a constraint on FDM model parameters. We show how to apply this analysis by examining two galaxies: the Milky Way and Andromeda. The mass of the halo of the Milky Way is known to be about $10^{12} M_\odot$, with some uncertainty related to modelling the dark matter mass distribution \cite{MWDM}; thus, the core-halo relation of eq. (\ref{corehalo}) predicts a central soliton of mass $M\approx 1.4\times10^{9}M_\odot\times(10^{-22}\text{ eV}/m)$. The latter corresponds to $N/N_c \approx 0.1$, which is independent of $m$. The central black hole of the Milky Way, Sagittarius A*, has a mass of $M_{BH} \approx 4\times10^{6} M_\odot$, inferred by observing orbits of stars very near to the galaxy center \cite{MWBH}. Given these two inputs, we represent the Milky Way by the 5-pointed star in Figure \ref{fig:BHplot}; because it is well to the left of the lines, we conclude that the central soliton of the Milky Way is safe from collapse for the FDM axion masses considered here.\footnote{Some have even claimed that velocity measurements in the central bulge of the Milky Way actually suggest the existence of a condensate \cite{SchiveEvidence}, though other analyses suggest tension \cite{BarBlum}.}

Next we analyze Andromeda, also known as M31. Andromeda is only slightly heavier than the Milky Way, less than a factor of $2$ more massive according to recent measurements \cite{AndromedaDM}; the corresponding central soliton predicted by eq. (\ref{corehalo}) is very similar to that of the Milky Way. On the other hand, its central black hole is much heavier, with a mass of roughly $M\approx2\times10^{8}M_\odot$ \cite{AndromedaBH}. We represent Andromeda by the 6-pointed star in Figure \ref{fig:BHplot}. Had the black hole or soliton of M31 been only a factor of a few heavier, this point would be to the right of the dash-dotted line, and the stability of M31 would be in tension with a FDM interpretation with $m\sim 10^{-20}$ eV. These simple examples motivate a more thorough study of currently-known supermassive black holes, to further test the FDM paradigm in this way. We leave such a study for a future work.

\subsection{UCDs as FDM Remnants}
It has been observed that some large galaxies contain so-called Ultracompact Dwarfs (UCDs) \cite{UCD1,UCD2}. These very compact sub-galaxies, which typically have a large mass-to-light ratio, have been interpreted by some as representing the very large mass tail of the globular cluster mass distribution \cite{UCDGC}. On the other hand, they can plausibly be interpreted as subhalos which were tidally stripped in their host galaxy, leaving only their core intact \cite{UCDTidal1,UCDTidal2}. The latter is supported, for example, by the presence of supermassive black holes inside of the heaviest known UCDs \cite{UCDformation}

If UCDs are dark matter dominated, they might plausibly be interpreted as FDM condensates (plus some remaining stars) whose outer virialized layers have been tidally stripped. It is interesting, for example, that the two densest UCDs currently known (M59-UCD3 and M85-HCC1 \cite{UCDDensest}) have masses of $\Ocal(10^7-10^8) M_\odot$ and radii of $\Ocal(1-100)$ pc, both in the vicinity of the maximum mass (and corresponding minimum radius) of FDM condensates with $m\sim10^{-21}-10^{-20}$ eV. It is worth noting that others have pointed out that FDM substructure could account for so-called Ultrafaint Dwarf galaxies, including \cite{Calabrese}, who suggested that the data preferred relatively low masses in the range $m\sim 3.7-5.6\times10^{-22}$ eV, though these authors neglect self-interactions that could potentially be important in the condensate. The scales involved are suggestive and it is interesting to consider what such a scenario might predict.

A plausible scenario to explain the cause of the tidal stripping of UCDs is the passage of a small galaxy near, or through, another large galaxy; such a close passage would leave behind only the ”nucleus” of the small galaxy (which potentially includes a black hole), thus forming a UCD \cite{UCDformation}. One might wonder whether this process would also induce collapse of an FDM condensate, through gravitational effects. As a "worst case” scenario, we analyzed the process of an FDM condensate, corresponding to the small galaxy's "nucleus", passing directly through the bulge of a large (Milky Way-like) galaxy, which may have occurred in a UCD's past. This process is analogous, in the sense of gravitational effects, to an axion star passing into an ordinary star, considered in \cite{ELLSW}. We find that FDM condensates are extremely stable under circumstances like these, with even a very massive bulge contributing a negligible effect.

Of course, it is possible that the stripping process in question could destroy the FDM substructure by pulling it apart rather than causing collapse; this depends on the mass of the core as well as its orbital radius around the galaxy. In \cite{Du}, it was found that the strongest constraints arising from these considerations are on relatively low-mass FDM models, $m\sim10^{-22}$ eV; because our analysis is most sensitive to high-mass regions (see Figure \ref{fig:BHplot}), this indicates the complementarity of our approach to previous analyses.

Just as very massive galaxies like the Milky Way are known to contain supermassive black holes, so too do some UCDs; for example, two of the densest UCDs ever discovered have central black holes which constitute greater than $10\%$ of their total mass \cite{UCDSMB}, and there is reason to expect that up to $80\%$ of all UCDs contain black holes \cite{UCDformation}. Discovery of new UCDs is accelerating, as data about galactic substructures becomes increasingly precise (a striking example being Gaia \cite{Gaia}). As we explained in the previous section, even a single galaxy or subgalaxy with a central black hole that places it to the right side of the curve in Figure \ref{fig:BHplot} would provide a significant constraint on the model. This could be relevant for UCDs as well.

Before leaving this section, we should point out that galaxy collisions can also lead directly to collapse of FDM condensate cores. This scenario is qualitatively similar to that considered in \cite{ELLSW}, where small condensates populate a galaxy and occasionally collide. 
One could in principle analyze these collisions given a particular model for halo mergers in cosmological history, but such an analysis is beyond the scope of this work. UCDs could collide with one another as well, but since the preponderance of UCDs is not currently known with great confidence, we do not attempt any further investigation of this scenario at this time.

\section{Analysis of Collapse Process} \label{sec:Decay}

\subsection{Collapse Times}
We will now apply the formalism of Section \ref{sec:Analytic} to the case of a condensate comprised of dark matter axions in FDM models. Although the model parameters can differ, we will take as our benchmark point an FDM axion with mass $m = 10^{-22}$ eV and decay constant $f = 10^{16}$ GeV, but we allow the former to also vary by a few orders of magnitude. First, recall that in previous works \cite{ELSW,ELSW2}, we expanded the energy only to the first order of $\delta$, which for the QCD axion was of $\Ocal(10^{-14})$. For our benchmark FDM parameters, one finds $\delta = \mathcal{O} \left(10^{-6}\right)$, and so while corrections will be much larger in this case, an expansion in $\d\ll1$ is still extremely appropriate.

Some physical parameters can be calculated in FDM models by a simple rescaling of $m$ and $f$. For example, the physical maximum mass and corresponding radius in dimensionless units are model independent (e.g. $n_c^{(QCD)} = n_c^{(FDM)}$); for our benchmark parameters we have $M_{c} \sim 10^{40}$ kg and $R_{c} \sim 10^{15}$ km (respectively), which can be obtained from the expressions in Section \ref{sec:Analytic}. On the other hand, because $\rho_D$ depends explicitly on $\d$, we find $\rho_D^{(FDM)} \approx 10^{-3}$ at $n=n_c$, even though $\rho_D^{(QCD)} = \Ocal(10^{-5})$ at the same $n$ \cite{ELSW}. This point corresponds to a physical radius of $R_D \sim10^{13}$ km in FDM, whereas the corresponding Schwarzschild radius is about $R_S \sim 10^{11}$ km.

\begin{figure}
 \includegraphics[scale=.55]{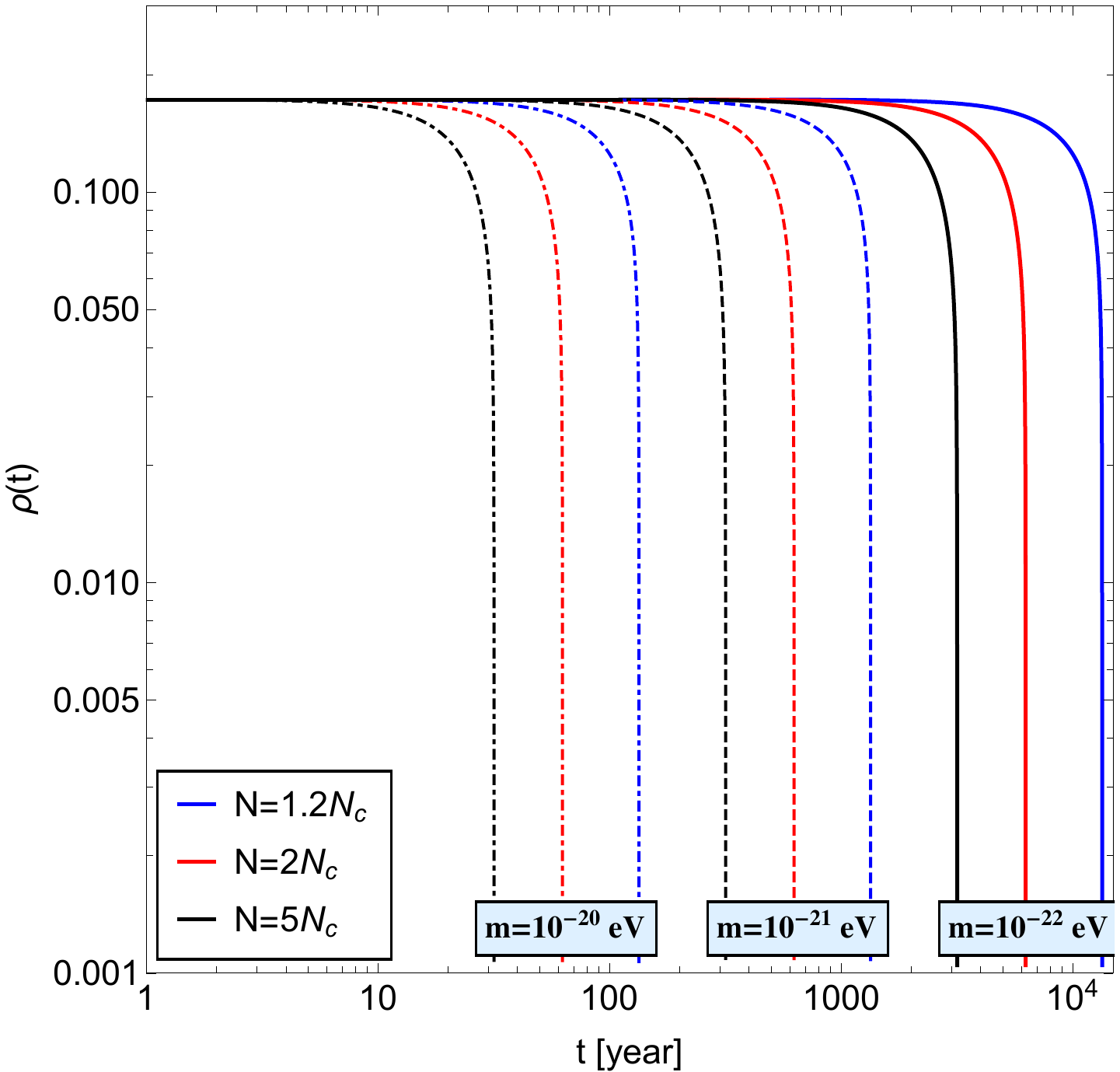}
 \includegraphics[scale=.55]{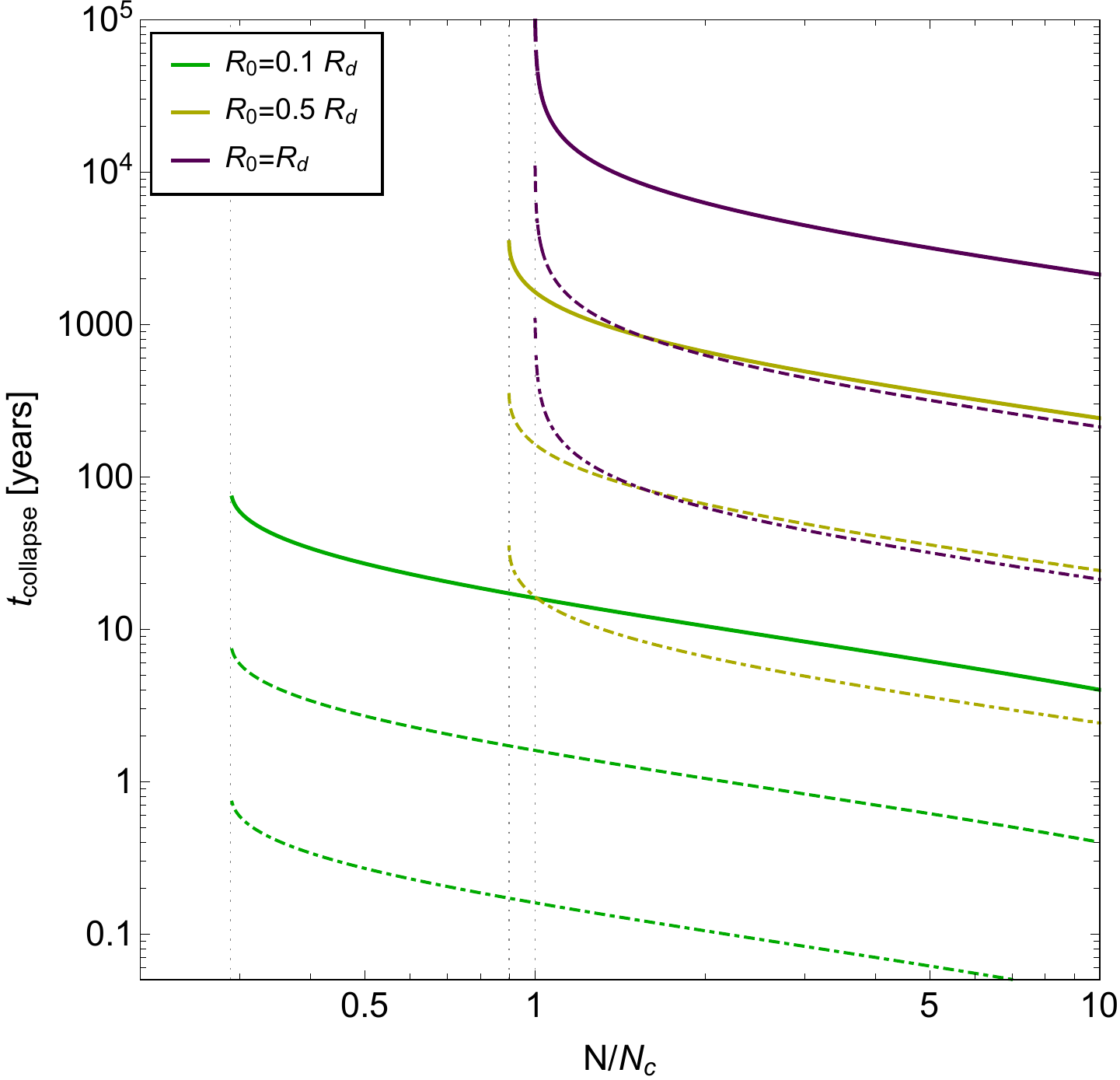}%{tCollapse_FDM}
 \caption{Collapse times for FDM condensates. In both plots, the particle mass $m$ is $10^{-22}$ eV (for thick lines), $10^{-21}$ eV (dashed), and $10^{-20}$ eV (dot-dashed), while the decay constant is $f = 10^{16}$ GeV.} 
 
 \underline{Left:} Rescaled radius $\r$ as a function of time, during collapse from $\r_d$ and $\r_D$. The particle number $N$ in the plot is $1.2\,N_c$ (for blue lines), $2\,N_c$ (red), and $5\,N_c$ (black).
 
 \underline{Right:} Total collapse time, calculated using eq. (\ref{eq:CollapseTime}) for FDM condensate as a function of particle number $N$. The initial radius $R_0$ in the collapse is $0.1R_d$ (for green lines), $0.5R_d$ (yellow), and $R_d$ (purple).
 \label{fig:tCollapse}
\end{figure}

The dependence of the  dimensionless radius $\r_D$ on $\d$ leads to changes in the collapse time in addition to the overall rescaling of $m$ and $f$ in the prefactor of eq. (\ref{eq:CollapseTime}). In the left panel of Figure \ref{fig:tCollapse}, we show the changing radius during collapse, for different choices of initial particle number $N$ and particle mass $m$. For example, when $N=1.2\,N_c$, the total collapse time is roughly $100$ years for $m=10^{-20}$ eV, $10^3$ years for $m=10^{-21}$ eV, and $10^4$ years for $m=10^{-22}$ eV. Recall that the collapse time is dominated by the more flat part of the potential, near $\rho_d$, whereas the collapse proceeds extremely fast during the final approach towards $\rho_D$ \cite{ELSW}. The radius is thus effectively flat until the very end of the collapse process.

In the right panel, we show the collapse times for the benchmark FDM model as a function of the initial particle number, for different choices of starting radii. $R_0 \neq R_d$ is relevant if the collapse were stimulated by some external perturbation of the radius. It is clear in either panel of the figure that a FDM condensate can be "collapsing" for several thousand years, and for the vast majority of this time there is no important change to its radius.

\subsection{Decay}
Collapsing condensates are subject to stimulated emission of relativistic particles during the final stages of their collapse \cite{ELSW,Levkov,Helfer,Michel}. This decay process has been treated using a relativistic field theoretic formalism in \cite{ESW,EMSW}, as well as in a nonrelativistic effective field theory \cite{MTY}, and the predictions have been well-confirmed in simulations of collapsing condensates \cite{Levkov,Helfer,Michel}. In the relativistic formalism, one finds that the decay rate depends on the parameter
\begin{equation} \label{eq:Delta}
 \D = \sqrt{1 - \frac{\m_0{}^2}{m^2}}
\end{equation}
in a straightforward way, where $\m_0$ is the eigenenergy (chemical potential) of the axion star. Using the variational analysis of Section \ref{sec:Analytic}, we can estimate the chemical potential as a function of the rescaled radius $\rho$ (see e.g. \cite{EbyThesis}):
\begin{equation} \label{eq:chemPot}
 \frac{\m_0(\r)}{m} = 1 + \d\left[\frac{3}{4}\frac{1}{\rho^2}
	- \frac{2}{\sqrt{2\pi}}\frac{n}{\rho} 
	    - \frac{1}{16\pi\sqrt{2\pi}}\frac{n}{\rho^3}
	    +\frac{3\,\delta}{864\pi^3\sqrt{3}}\frac{n^2}{\rho^6}\right],
\end{equation}
which we compute to next to leading order in the self-interaction, as before. If we treat this quantity as directly varying with $\r$, we can calculate $\D$ directly. Finally, we can use the result for the $3\,a_c \to a_f$ decay rate of axion stars, which we calculated in \cite{ESW}, and later applied to FDM scenarios in \cite{EMSW}:
\begin{equation} \label{Gamma}
 \Gamma_3(\r) = \frac{2\,\pi\,f^2}{\sqrt{8}\,m}
	\Big[\frac{32\pi \,y_I}{3\Delta(\r)}
	  \exp\Big(-\frac{\sqrt{8}y_I}{\Delta(\r)}\Big)\Big]^2,
\end{equation}
where $y_I \approx 0.603156$. Putting all of this together, we can track the decay rate for $3\,a_c \to a_f$ during the collapse process. This will allow us to approximate the relevant timescales for collapsing FDM condensates.

In the case of the QCD axion, as considered in \cite{ELSW}, the dilute and dense minima occur at $\rho = \Ocal(1)$ and $\rho \sim 10^{-5}$, respectively. The total collapse takes a few minutes, but the biggest changes in $\rho$ occur only in the last fraction of a second. As a result, the decay rate (\ref{Gamma}) is extremely small until $\rho \lesssim 10^{-4}$, which (comparing with the results of \cite{ELSW}) does not occur until the last fraction of a second of the collapse. The timestep between an extremely negligible rate (when $\rho < 10^{-4}$) to an extremely large rate (when $\rho \sim 10^{-5}$) was smaller than $10^{-7}$ sec. As a result, in \cite{ELSW}, we approximated the decay as though it turns on instantaneously, which is a very good approximation.

\begin{figure} [t]
 \includegraphics[scale=.39]{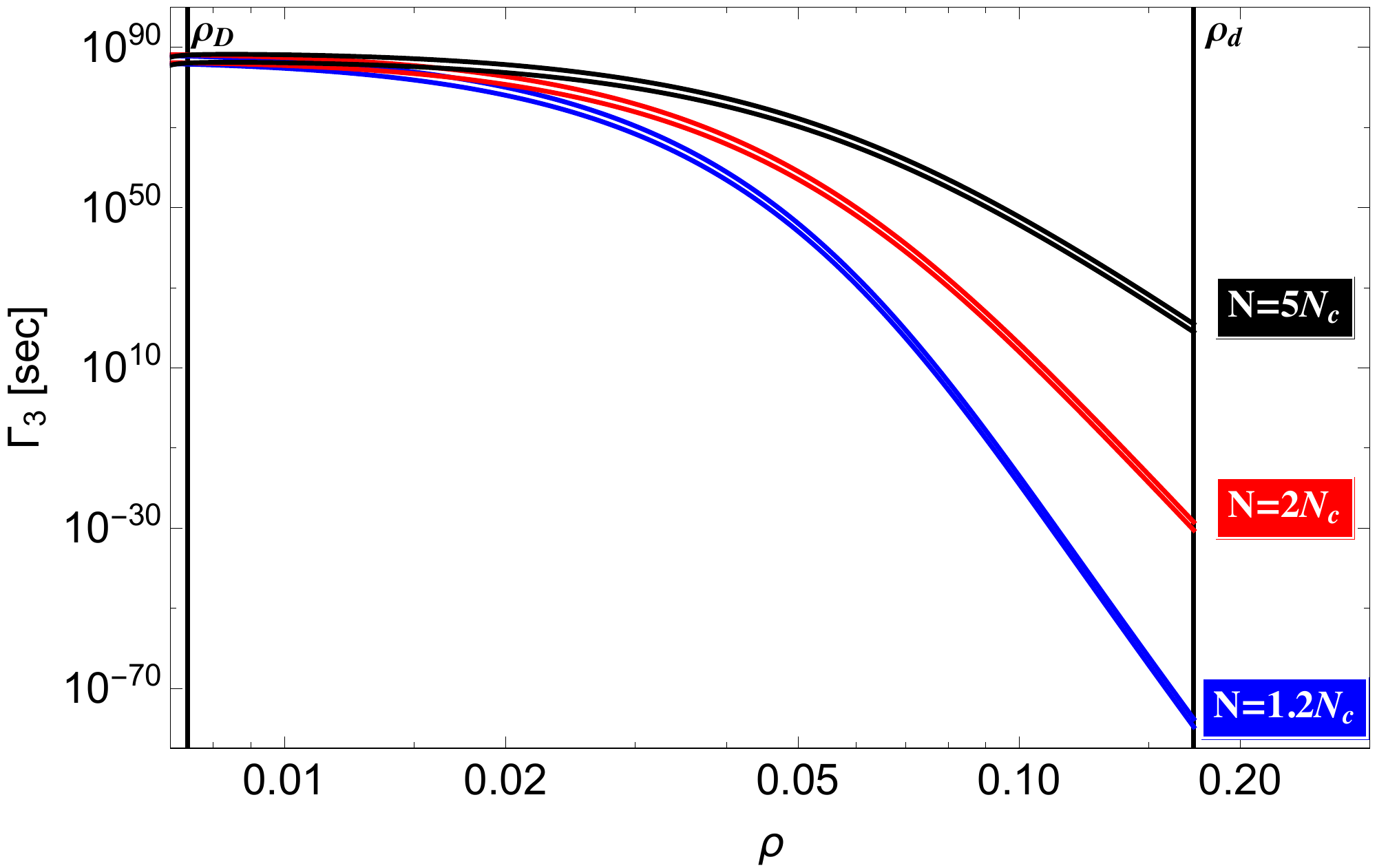}
 \includegraphics[scale=.39]{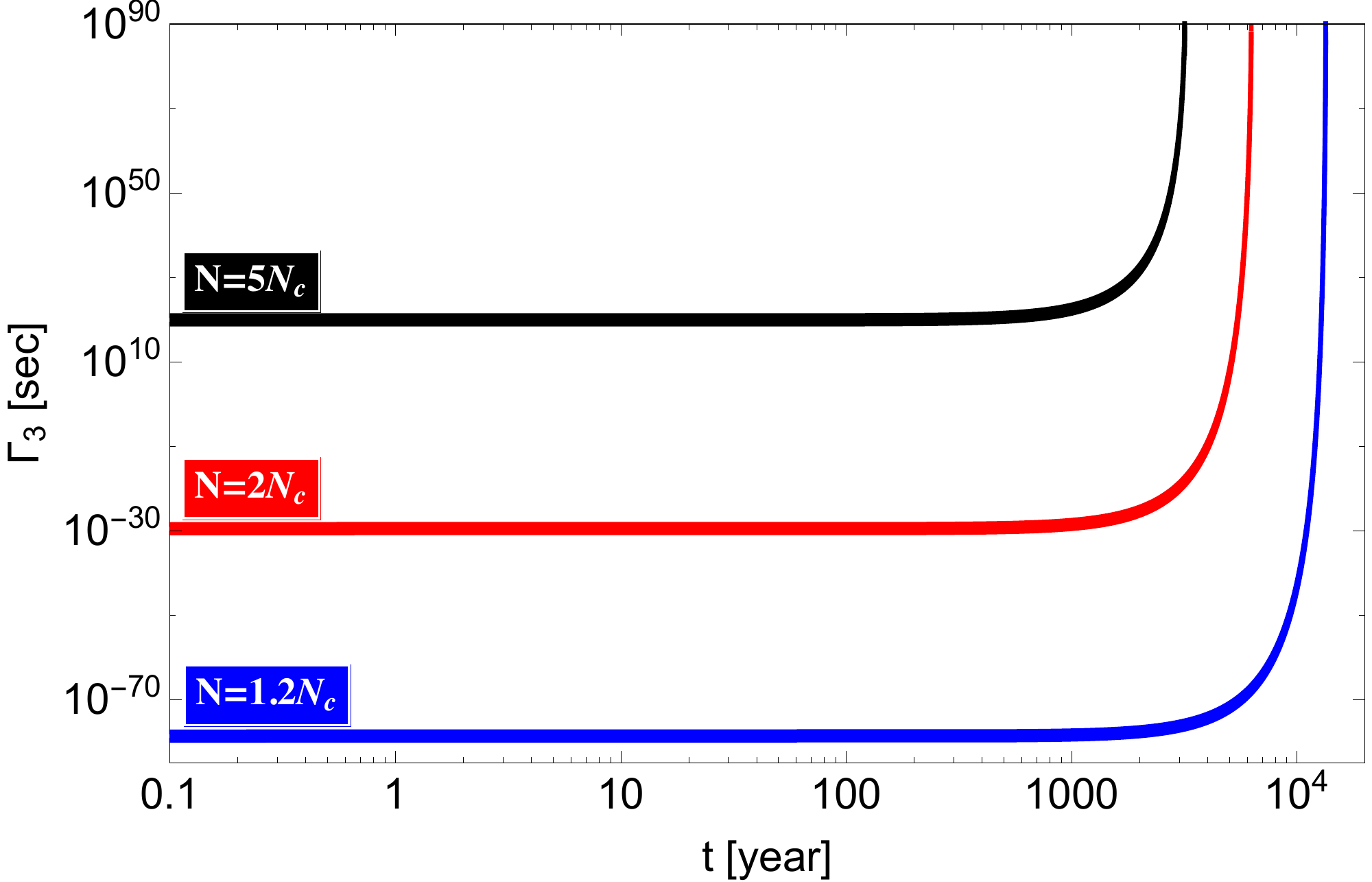}
 \caption{The decay rate of (\ref{Gamma}) as a function of the rescaled radius $\rho$ (left) and the collapse duration (right), for the FDM axion and a choice different choices of particle mass $N=1.2N_c$, $2N_c$, and $5N_c$ (blue, red, and black curves, respectively). The thickness of the curves shows the change upon varying the axion mass in the range of $m\in\{10^{-20}-10^{-22}\}$ eV.}
 \label{fig:Decay}
\end{figure}

For the FDM case with $f=10^{16}$ GeV, there are some differences. The rescaled dilute radius is still $\rho = \Ocal(1)$ but the dense one moves to $\rho \sim 0.005$, much larger than the QCD case (see Figure \ref{fig:Decay}, left panel). The decay rate still starts out very relatively small, but during the collapse it turns on not in a fraction of a second, but over the course of many years (see Figure \ref{fig:Decay}, right panel). We have shown the curves for supercritical particle numbers $N=\{1.2,2,5\}\times N_c$ to show the parametric dependence of the decay rate. In the third case, it is interesting that the decay rate is as high as $10^{20}$ axions/sec, even before collapse. However, it is not clear by what mechanism an FDM condensate might attain a particle number as high as $5N_c$, so this remains merely a curiosity.

In Figure \ref{fig:Decay}, we can see that once $\rho \lesssim 0.02$, the emission rate is $\G_3 = \Ocal(10^{80})$ axions/sec, which for axions of mass $m = 10^{-22}$ eV gives
\begin{equation}
 \text{Mass Loss Rate} \sim \G_3\times m \sim 1\,\frac{M_\odot}{sec}.
\end{equation}
Near this point, the decay process dominates the dynamics of the collapse, leading to an important backreaction.  This is seen in the numerical simulations of \cite{Levkov,Helfer,Michel}, which suggest that as rapid emission begins, the condensate "bounces" against a hard-core repulsion in the core of the star, possibly returning to a stable dilute configuration after an $\Ocal(few)$ number of such bounces. Of course, in our formalism at fixed $N$ we are not sensitive to this backreaction or  "bounce" effect. It is intriguing, however, that even during a single collapse epoch, a FDM condensate would rapidly emit relativistic axions over a timespan of many hundreds or thousands of years. It would be interesting to investigate further possible consequences of this result in realistic galactic scenarios.

We should also note that, if a FDM condensate surrounds a black hole, the latter will slowly accrete mass, though the rate is suppressed when the condensate radius is very large \cite{Witten,BarBlum}. During collapse, however, the rate can be enhanced by additional time dependent factors in the wavefunction, so that the black hole might see a large increase in its mass in the final state. We leave a full investigation of this effect to future work.

\section{Conclusions} \label{sec:Conclusions}

In this work, we have studied the stability of Fuzzy Dark Matter (FDM) halos made of ultralight axion-like particles. Our analysis applies specifically to the soliton-like Bose-condensed structure of axions, which could be either a free-standing configuration or a central region of a large galaxy surrounded by uncondensed gas of axion particles populating the outer regions of the galaxy. These weakly-bound structures in dark matter halos can become structurally unstable through self-collisions, or through collisions with other astrophysical objects; in any case, instability leads to collapse towards a denser configuration. The collapse times range from hundreds to thousands of years, depending on the mass of the axion particle.

Although the collapse formally takes many years, a large fraction of this time there is no significant change in the radius of the condensate. During the final approach to the dense configuration, relativistic effects begin to dominate as the condensate rapidly emits high-energy axions. For FDM parameters, the turn-on of this decay process can be slow, lasting hundreds of years, before the backreaction becomes sufficiently strong to lead to a hard-core repulsion that counteracts the collapse \cite{Levkov,Helfer,Michel}.

We have investigated the interpretation of Ultracompact Dwarf (UCD) galaxies as being primarily composed of FDM condensate cores. The mass and radius scale of these UCDs is suggestive of a connection to the typical condensate mass and size in FDM.  We have argued that an interaction in the host galaxy which gives rise to tidal stripping of the UCD would not significantly affect its stability.

A more probable scenario is an interaction between an FDM condensate and a large black hole. Central black holes in galaxies are much smaller in dimension than a typical FDM halo, which has an extent of hundreds of light years. We have seen that if such a black hole enters the region encompassed by the FDM halo, they can cause a structural instability leading to collapse and decay. In particular, supermassive black holes in large galaxies could constrain FDM models, since the central soliton's stability is not guaranteed in the presence of external potentials; we illustrated the power of this idea using the Milky Way and Andromeda galaxies as simple examples. One could plausibly constrain a potential FDM component to UCDs, if they are observed to contain sufficiently massive supermassive black holes; some UCDs have already been observed to contain such \cite{UCDSMB}. This could be an important consideration for determining the structure of UCDs in the future. We leave a more detailed analysis of this interpretation to a future publication.

\section*{Acknowledgements}
 
We thank P. Argyres, N. Bar, K. Blum, R. Gass, M. Ma, S. Schon, L. Street, M. Takimoto, and C. Vaz for fruitful discussions. The work of J.E. was supported by the Zuckerman STEM Leadership Program. M.L. thanks the Barry Goldwater Scholarship and Excellence in Education Foundation for scholarship support, and the Department of Physics at the University of Cincinnati for financial support in the form of a Violet Diller Fellowship.

 \appendix
 \section{Tunneling From A Metastable Condensate} \label{AppA}
 \renewcommand{\theequation}{A.\arabic{equation}}
  
It was shown many years ago that the tunneling rate of a metastable condensate has the form \cite{Stoof}
\begin{equation}
 \Gamma = P \exp\left[-2\,N\,S_E\right]
\end{equation}
where $S_E$ is the Euclidian action and $P$ is a prefactor which can be evaluated using a single-instanton formalism. In this appendix we will evaluate this rate using the leading-order axion potential. The result presented here is mostly based on the work of \cite{Stoof,Arovas}. It overlaps somewhat with \cite{Marquardt} who also investigated $1/r$ potentials, though in the details their method of analysis is very different.

The tunneling rate is related to the solutions to the equation of motion defined by $S_E$.  It is not hard to show that, for a potential $E(R)$, the equation of motion (in Euclidian time $\tau$) is that of a classical particle moving through a potential $-E(R)$; that is,
\begin{equation}
 \frac{d^2 R(\tau)}{d\tau^2} = - \frac{1}{N}\ddx{(-E(R))}{R}
\end{equation}
One solution to this equation is the trivial one: The ``particle'' sits still at $R_0$.  The 
non-trivial solution is the so-called ``bounce'', where asymptotically 
${R(\tau\rightarrow\pm\infty)=R_d}$ but at some finite time $\tau_1$, the particle bounces between the local minimum $R_d$ and the equal-energy point $R_1$ on the other side of the barrier. (These two points are connected by the dashed lines in Figure \ref{fig:Energy}.)  Such trajectories $R_b(\tau)$ are known as instantons.

%\red{Cite Masahiro / Ryosuke paper on bounce actions, 1707.01099? Maybe also Espinosa 1805.03680}

We begin with $S_E$, which is a WKB integral over the energy barrier of the classical momentum per particle, $p(R) = \sqrt{2\,m\,\D E/N}$ \cite{Stoof}:
\begin{align} \label{eq:SE}
 S_E &= \int_{R_1}^{R_d}{dR\sqrt{\frac{2\,m}{N}\,\left[E(R)  - E(R_d)\right]}} \nn \\
 	&= \int_{R_1}^{R_d}{dR\sqrt{2(m^2\,\d)\,\left[e(\r)  - e(\r_d)\right]}} \nn \\
 	&= \sqrt{2}\int_{\rho_1}^{\rho_d}{d\rho\sqrt{e(\rho) - e(\rho_d)}}
 \end{align}
Of course, in the range $\rho_1 \leq \r \leq \r_d$, the integrand is real (see Figure \ref{fig:Energy}). Now, the prefactor for the metastable tunnelling rate generically has the form \cite{Stoof,Coleman,Milnikov}
\begin{equation} \label{Prefactor}
 P = \sqrt{\frac{S_E}{2\pi}}\Bigg|\frac{\det^\prime\Big(-\partial_\tau^2 
				      + \frac{E''(R)}{m\,N}\Big)}
				     {\det \Big(-\partial_\tau^2 + \omega_0^2 \Big)} \Bigg|
	= L \,\omega_0 \sqrt{\frac{N\,m\,\omega_0}{\pi}}
\end{equation}
which depends both on the curvature of the potential at the local minimum
\begin{align}
 \omega_0^2 &\equiv \frac{1}{m\,N}\frac{\partial^2 E}{\partial R^2}\Big|_{R_d} \nn \\
 			&= m^2\,\frac{f^3}{M_P{}^3}\frac{\partial^2 e(\r)}{\partial\r^2}\Big|_{\r_d}
\end{align}
as well as a coefficient $L$, defined by the asymptotic form of the ``bounce'' solutions \cite{Milnikov}
\begin{equation}
 R_b(\tau) \xrightarrow{\tau \rightarrow \pm \infty} R_d - L e^{-\omega_0 |\tau|}.
\end{equation}
This is appropriate because of the approximately quadratic form of the inverted potential near $R_d$.  
However, we also know that the bounce solution oscillates quickly to $R_1$ at some time $\tau_1$, which 
by time-translation invariance might as well be set to $\tau_1 = 0$; thus, we also require
\begin{equation}
 R_b(\tau=0) = R_1.
\end{equation}

Now we write the classical momentum as $p = m\,\partial R/\partial\tau$ to find an expression for the $R$-dependence 
of $\tau$:
\begin{align} \label{tau}
 \omega_0\,\tau &= \omega_0 m \int_{R_d}^{R(\tau)}{\frac{dR^\prime}{p(R^\prime(\tau))}} \nonumber \\
      &= \int_{R_d}^{R}{dR^\prime \Big[\frac{m\omega_0}{p(R^\prime)} 
			- \frac{1}{R_d-R^\prime} \Big]} 
			+ \ln\Big(\frac{R_d-R(\tau)}{R_d-R^\prime}\Big)\Bigg|_{R^\prime = R_d}
			+ \omega_0\,c \nonumber \\
      &= \Ical(R) + \ln\Big(\frac{R_d-R(\tau)}{R_d-R^\prime}\Big)\Bigg|_{R^\prime = R_d} + \omega_0\,c
\end{align}
where we have defined the first integral as $\Ical(R)$. Since $p(R) = 0$ at $R = R_d$ and is proportional to $R$ nearby, we have regularized the integral by subtracting this log-divergent piece. The third term is an integration constant, which we determine by using 
the fact that $R_b(\tau=0) = R_1$:
\begin{equation}
 \omega_0\,c = -\Ical(R_1) - \ln\Big(\frac{R_d-R(\tau)}{R_d-R^\prime}\Big)\Bigg|_{R^\prime = R_d},
\end{equation}
cancelling the log-divergent piece in eq. (\ref{tau}). Thus, for $\tau < 0$ (the first half of the ``bounce''),
\begin{align}
 \omega_0\,\tau = -\omega_0\,|\tau|= \Ical(R) - \Ical(R_1) + \ln\Big(\frac{R_d-R(\tau)}{R_d-R_1}\Big).
\end{align}
At large $|\tau|$, $\Ical(R) \approx \Ical(R_d) \ll \Ical(R_1)$ (since the integrand is positive-definite), and so we neglect this term. Rearranging the equation gives the result
\begin{equation}
 R_d - R(\tau) = \Big[(R_d - R_1)e^{\Ical(R_1)} \Big] e^{-\omega_0 |\tau|}
\end{equation}
We identify the expression in square brackets as the coefficient $L$, i.e.
\begin{equation}
 L =  (R_d - R_1)e^{\Ical(R_1)}
\end{equation}
which matches the result of \cite{Milnikov}.  We can also rewrite $I(R_1)$ in terms of 
dimensionless quantities,
\begin{align} \label{eq:intI}
 \Ical \equiv \Ical(R_1) &= \int_{R_d}^{R_1}{dR \Big[\frac{m\omega_0}{p(R)}-\frac{1}{R_d-R} \Big]} 
						  \nn \\
	&= \frac{1}{\sqrt{2}} \int_{\r_d}^{\r_1} d\r\,\frac{\sqrt{e''(\r)}}{\sqrt{e(\r) - e(\r_d)}}
			- \int_{\r_d}^{\r_1} d\r\,\frac{1}{\r_d - \r}
\end{align}

\begin{figure}
 \centering
 \includegraphics[scale=1]{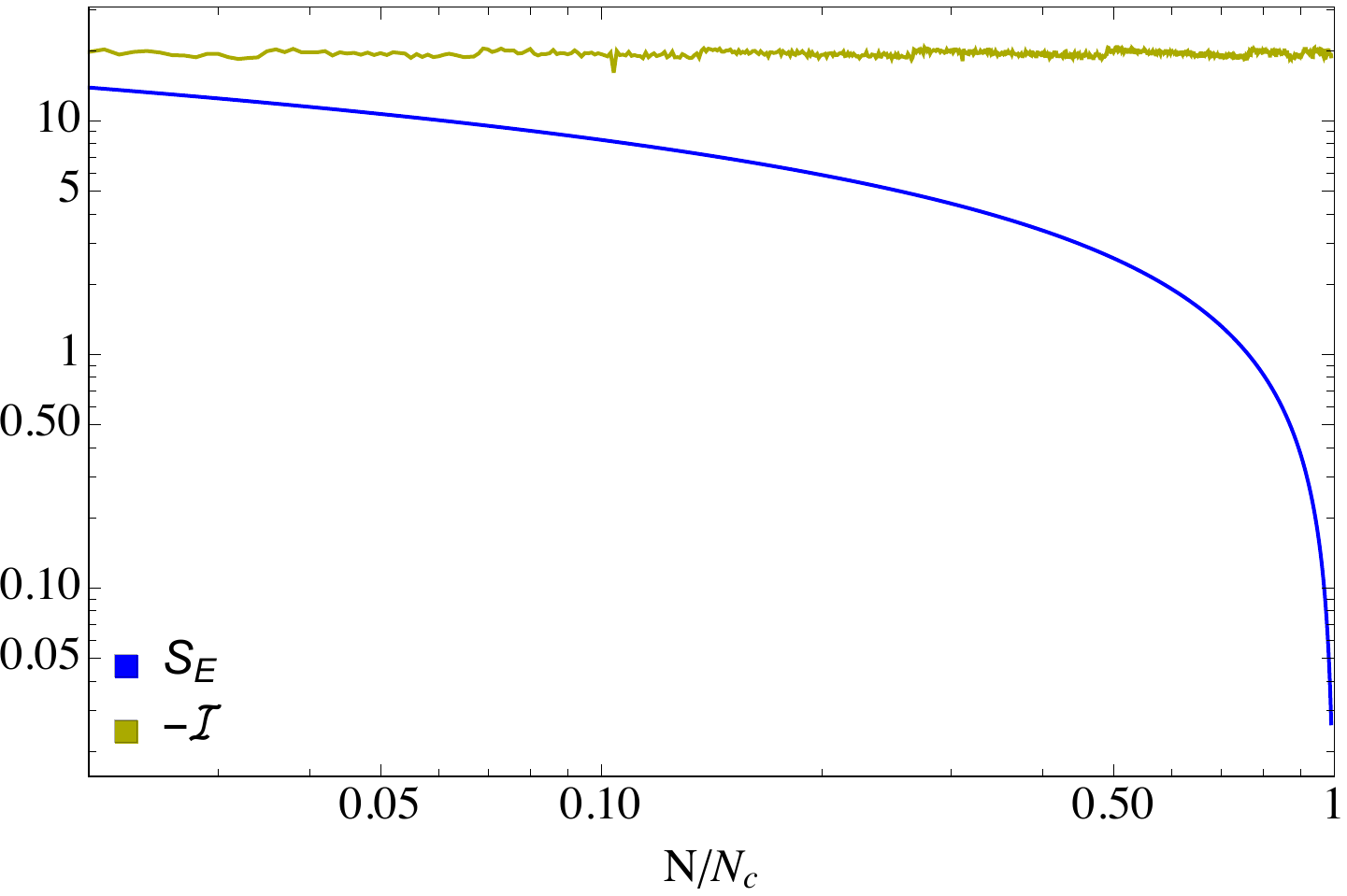}
 \caption{The numerical integration of $S_E$, eq. (\ref{eq:SE}), and $-\Ical$, eq. (\ref{eq:intI}) (inverted by a minus sign), over different values of $N$. As expected, $\Ical$ (yellow curve) is essentially independent of $N$ (scatter is due to numerical integration), whereas $S_E$ (blue curve) falls fast when $N$ is very close to $N_c$.}
 \label{fig:Tunnelling}
\end{figure}

The result for the tunnelling rate is
\begin{align}
 \G &= (R_d - R_1) \,\o_0\, \sqrt{\frac{N\,m\,\o_0}{\pi}} \exp\left(\Ical - 2\,N\,S_E\right) \nn \\
 	&= \frac{m\sqrt{N}}{\sqrt{\pi}}\left(\frac{f}{M_P}\right)^{5/4} 
			\left(\r_d - \r_1\right)\,\left[e''(\rho)\right]^{3/2}\exp{\left(\Ical - 2\,N\,S_E\right)}.
\end{align}
Note that the exponential $\exp\left(\Ical - 2\,N\,S_E\right) \approx \exp\left(- 2\,N\,S_E\right) \lll 1$ when $N\gg1$ (appropriate in the condensate). This term dominates the decay rate. We checked numerically that $S_E$ is not sufficiently small to counteract the largeness of $N$ unless the particle number is extremely close to $N_c$. Similarly, $\Ical$ is never large enough to be relevant to the decay rate; the value of $\Ical \approx -19$ is independent of $N$. The results for both integrals are shown in Figure \ref{fig:Tunnelling}. Because the rate is suppressed by $\exp\left(-N\right)$, we conclude that the effect of tunnelling is negligible (as expected).


\begin{thebibliography}{100}
 \bibitem{Weinberg} D. H. Weinberg et al., "Cold dark matter: Controversies on small scales." Proceedings of the National Academy of Sciences 112 (2015) 12249-12255. arXiv: 1306.0913
 \bibitem{NoCuspCore} A. Genina et al., "The core–cusp problem: a matter of perspective." Monthly Notices of the Royal Astronomical Society 474, Issue 1 (2018) 1398–1411. arXiv: 1707.06303
 \bibitem{NoMissSat} S. Y. Kim, A. H. G. Peter, and J. R. Hargis, "There is No Missing Satellites Problem." arXiv: 1711.06267
 \bibitem{NoFDMCuspCore} H. Deng, M. P. Hertzberg, M. Hossein Namjoo, and A. Masoumi, "Can Light Dark Matter Solve the Core-Cusp Problem?" arXiv: 1804.05921

 \bibitem{Kaup} D.J. Kaup, ``Klein-Gordon Geon.'' Phys. Rev 172 (1968) 1331.
\bibitem{RB} R. Ruffini and S. Bonazzola, ``Systems of Self-Gravitating Particles in General Relativity and the Concept of an Equation of State.'' Phys. Rev. 187 (1969) 1767.
\bibitem{BreitGuptaZaks}  J.D. Breit, S. Gupta, and A. Zaks, ``Cold Bose stars." Phys. Lett. B {\bf140} (1984) 329–332.
\bibitem{CSW} M. Colpi, S. L. Shapiro, and I. Wasserman, ``Boson Stars: Gravitational Equilibria of Self-Interacting Scalar Fields.'' Phys. Rev. Lett. {\bf 57} (1986) 2485.
\bibitem{SS} E. Seidel and W-M Suen, ``Dynamical evolution of boson stars: Perturbing the ground state.'' Phys. Rev D. {\bf42} (1990) 384.

\bibitem{Friedberg} R. Friedberg and T.D. Lee and Y. Peng, ``Scalar Soliton Stars and Black Holes.'' Phys. Rev. D {\bf35} (1987) 3658.
\bibitem{SS2} E. Seidel and W-M Suen, ``Oscillating soliton stars.'' Phys. Rev. Lett. 66 (1991) 1659.
\bibitem{Liddel} A. Liddel and M. Madsen, ``The Structure and Formation of Boson Stars.'' Int. Journal Mod. Phys. D1 (1992) 101.
\bibitem{Lee} T. D. Lee and Y. Peng, ``Nontopological solitons.'' Phys. Rep. 221, 251 (1992).

\bibitem{Khlopov1} M. Y. Khlopov, B. A. Malomed, and Y. B. Zeldovich, ''Gravitational Instability of Scalar Field and Primordial Black Holes.``
Mon. Not. Roy. Astr. Soc., V. 215, PP.575-589 (1985).

\bibitem{Khlopov2}Z. G. Berezhiani, A. S. Sakharov, and M. Yu. Khlopov,
''Primordial background of cosmological axions.`` Yadernaya Fizika (1992) V.
55, PP. 1918-1933.
[English translation: Sov.J.Nucl.Phys. V.55, PP.1063-1071 (1992)]

\bibitem{Khlopov3}A. S. Sakharov and M. Y. Khlopov,
''The nonhomogeneity problem for the primordial axion field.`` Yadernaya Fizika
(1994) V. 57, PP. 514- 516. 
[English translation: Phys.Atom.Nucl. V. 57, PP. 485-487 (1994)]

\bibitem{Khlopov4}A. S. Sakharov, D. D. Sokoloff, and M. Y. Khlopov,
''Large scale modulation of the distribution of coherent oscillations of a
primordial axion field in the Universe.`` Yadernaya Fizika (1996) V. 59, PP.
1050-1055. [English translation: Phys.Atom.Nucl. V. 59, PP. 1005-1010 (1996)]

\bibitem{Khlopov5}M. Y. Khlopov, A. S. Sakharov, and D. D. Sokoloff,
''The nonlinear modulation of the density distribution in standard axionic CDM
and its cosmological impact.`` Nucl.Phys. B (Proc. Suppl.) V. 72,
105-109 (1999)

  \bibitem{Iwazaki} A. Iwazaki, ``Axionic boson stars in magnetized conducting media.'' Phys. Rev. D {\bf60} (1999) 025001. arXiv: hep-ph/9901396
 \bibitem{SikivieYang} P. Sikivie and Q. Yang, ``Bose-Einstein Condensation of Dark Matter Axions.'' Phys. Rev. Lett. {\bf103} (2009) 111301. arXiv: 0901.1106
  \bibitem{ChavanisMR} P.H. Chavanis, ``Mass-radius relation of Newtonian self-gravitating Bose-Einstein condensates with short-range interactions: I. Analytical results.'' Phys. Rev. D {\bf84} (2011) 043531. arXiv: 1103.2050
 \bibitem{ChavanisMR2} P.H. Chavanis and L. Delfini, ``Mass-radius relation of Newtonian self-gravitating Bose-Einstein condensates with short-range interactions: II. Numerical results.'' Phys. Rev. D {\bf84}(2011) 043532. arXiv: 1103.2054.
  \bibitem{BB}  J. Barranco and A. Bernal, ``Self-gravitating system made of axions.`` Phys. Rev. D {\bf83} (2011) 043525. arXiv: 1001.1769
   \bibitem{BarrancoNS} J. Barranco, A. Carrillo Monteverde, D. Delepine, ``Can the dark matter halo be a collisionless ensemble of axion stars?'' Phys. Rev. D {\bf 87} (2013) 10, 103011. arXiv: 1212.2254
  \bibitem{TkachevFRB} I.I. Tkachev, ``Fast Radio Bursts and Axion Miniclusters.'' JETP Letters 101 (2015) 1. arXiv: 1411.3900 
  \bibitem{ESVW} J. Eby, P. Suranyi, C. Vaz, and L.C.R. Wijewardhana, ``Axion Stars in the Infrared Limit.'' JHEP 1503 (2015) 080.  arXiv:1412.3430 
  \bibitem{Guth} A. H. Guth, M. P. Hertzberg and C. Prescod-Weinstein, ''Do Dark Matter Axions Form a Condensate with Long-Range Correlation?'' Phys. Rev. D {\bf92} (2015) 103513 36. arXiv: 1412.5930
   \bibitem{Braaten} E. Braaten, A. Mohapatra, and H. Zhang, ``Dense Axion Stars.`` Phys. Rev. Lett. {\bf117} (2016) 121801. arXiv: 1512.00108
 \bibitem{ELSW} J. Eby, M. Leembruggen, P. Suranyi, and L.C.R. Wijewardhana, ``Collapse of Axion Stars,'' JHEP 1007 (2016) 066.  arXiv:1608.06911.
 \bibitem{ELSW2} J. Eby, M. Leembruggen, P. Suranyi, and L.C.R. Wijewardhana, ``QCD Axion Star Collapse with the Chiral Potential.'' JHEP 2017 (2017) 14. arXiv: 1702.05504
 \bibitem{WilczekASt} L. Visinelli, S. Baum, J. Redondo, K. Freese, and F. Wilczek, ``Dilute and dense axion stars.'' Phys. Lett. B {\bf777} (2018) 64-72. arXiv: 1710.08910
 \bibitem{ChavanisPT} P.H. Chavanis, ``Phase transitions between dilute and dense axion stars.'' arXiv: 1710.06268
  \bibitem{Kling1} F. Kling and A. Rajaraman, ``Towards an Analytic Construction of the Wavefunction of Boson Stars.'' Phys. Rev. D {\bf96} (2017) 044039. arXiv: 1706.04272
 \bibitem{Kling2} F. Kling and A. Rajaraman, ``On Profiles of Boson Stars with Self-Interactions.'' Phys. Rev. D {\bf97} (2018) 063012. arXiv: 1712.06539
\bibitem{Hertzberg1} E. D. Schiappacasse and M. P. Hertzberg, “Analysis of Dark Matter Axion Clumps with
Spherical Symmetry.” JCAP 1801 (2018) 037. Erratum: [JCAP 1803 (2018) no. 03, E01]
arXiv: 1710.04729
\bibitem{sarkar} S. Sarkar, C. Vaz and L.C.R. Wijewardhana, "Gravitationally Bound Bose Condensates with Rotation." To appear in Phys. Rev. D. arXiv: 1711.01219
\bibitem{Hertzberg2} M. P. Hertzberg and E. D. Schiappacasse, “Scalar Dark Matter Clumps with Angular Momentum,”
arXiv:1804.07255
\bibitem{Hertzberg3} M. P. Hertzberg and E. D. Schiappacasse, "Dark Matter Axion Clump Resonance of Photons." arXiv: 1805.00430
 
  \bibitem{HoganRees} C.J. Hogan and M.J. Rees, ``Axion Miniclusters.'' Phys. Lett. B {\bf205} (1988) 228-230.
 \bibitem{KolbTkachev} E.W. Kolb and I.I. Tkachev, ``Axion miniclusters and Bose stars.`` Phys. Rev. Lett. {\bf71} (1993) 3051-3054.
 \bibitem{TkachevSim} D.G. Levkov, A.G. Panin, and I.I. Tkachev, "Bose Condensation by Gravitational Interactions." arXiv: 1804.05857
 
 \bibitem{Nelson} A. Nelson and H. Xiao, "Axion Cosmology with Early Matter Domination." arXiv: 1807.07176
 \bibitem{Visinelli} L. Visinelli, "Axion Miniclusters in Modified Cosmological Histories." arXiv: 1808.01879
 
   \bibitem{ChavanisCollapse} P.H. Chavanis, ``Collapse of a self-gravitating Bose-Einstein condensate with attractive self-interaction.'' Phys. Rev. D {\bf94} (2016) 083007. arXiv: 1604.05904
   
    \bibitem{Bosenova} E.A. Donley, N.R. Claussen, S.L. Cornish, J.L. Roberts, E.A. Cornell, and C.E. Wieman, ``Dynamics of collapsing and exploding Bose-Einstein condensates.'' Nature 412, 295 (2001). arXiv: cond-mat/0105019
 
 \bibitem{ESW} J. Eby, P. Suranyi, and L.C.R. Wijewardhana, ``The lifetime of axion stars." Mod.\ Phys.\ Lett.\ {\bf A31} (2016) no.15, 1650090. arXiv: 1512.01709
 \bibitem{EMSW} J. Eby, M. Ma, P. Suranyi, and L.C.R. Wijewardhana, ”Decay of Ultralight Axion Condensates.“ JHEP 01 (2018) 066. arXiv: 1705.05385
   \bibitem{Levkov} D.G. Levkov, A.G. Panin, and I.I. Tkachev, ``Relativistic Axions from Collapsing Bose Stars.'' Phys. Rev. Lett. {\bf118} (2016) 011301. arXiv: 1609.03611
\bibitem{Helfer} T. Helfer, D. J. E. Marsh, K. Clough, M. Fairbairn, E. A. Lim, and R. Becerril, ``Black Hole Formation from Axion Stars.'' JCAP 03 (2017) 055. arXiv: 1609.04724
\bibitem{Michel} F. Michel and I. G. Moss, "Relativistic collapse of axion stars." arXiv: 1802.10085
 \bibitem{Hertzberg} M. P. Hertzberg, "Quantum radiation of oscillons." Phys. Rev. D {\bf82} (2010) 045022. arXiv: 1003.3459
  \bibitem{MTY} K. Mukaida, M. Takimoto, and M. Yamada, ``On Longevity of I-ball/Oscillon.'' JHEP 2017 (2017) 122. arXiv: 1612.07750.
  \bibitem{EMTWY} J. Eby, K. Mukaida, M. Takimoto, L.C.R. Wijewardhana, and M. Yamada, "Classical Nonrelativistic Effective Field Theory and the Role of Gravitational Interactions." arXiv: 1807.09795
  
  \bibitem{Cotner} E. Cotner, ``Collisional interactions between self-interacting non-relativistic boson stars: effective potential analysis and numerical simulations.'' Phys. Rev. D {\bf94} (2016) 063503. arXiv: 1608.00547
 \bibitem{ELLSW} J. Eby, M. Leembruggen, J. Leeney, P. Suranyi, and L.C.R. Wijewardhana, ``Collisions of Dark Matter Axion Stars with Astrophysical Sources.'' JHEP 04 (2017) 99. arXiv: 1701.01476
 
 \bibitem{Turner} M. S. Turner, ``Coherent scalar field oscillations in an expanding universe.'' Phys. Rev. D {\bf28} (1983) 1243-1247.
 \bibitem{Press} W. H. Press, B. S. Ryden, and D. N. Spergel, ``Single mechanism for generating large-scale structure and providing dark missing matter.'' Phys. Rev. Lett. {\bf64} (1990) 1084-1087.
 \bibitem{Sin} S.-J. Sin, ``Late time cosmological phase transition and galactic halo as Bose liquid.`` Phys. Rev. D {\bf50} (1994) 3650-3654. arXiv: hep-ph/9205208.
  \bibitem{Hu} W. Hu, R. Barkana, and A. Gruzinov, ''Cold and fuzzy dark matter.`` Phys. Rev. Lett. {\bf85} (2000) 1158-1161. arXiv: astroph/0003365.
 \bibitem{Goodman} J. Goodman, ''Repulsive dark matter.`` New Astron. 5, 103-107 (2000). arXiv: astro-ph/0003018
 \bibitem{Peebles} P. J. E. Peebles, ''Fluid dark matter.'' Astrophys. J. 534, L127-L129 (2000). arXiv: astro-ph/0002495
 \bibitem{Amendola} L. Amendola and R. Barbieri, ``Dark matter from an ultra-light pseudo-Goldstone-boson.'' Phys. Lett. B {\bf642} (2006) 192-196. arXiv: hep-ph/0509257
 \bibitem{Shapiro} B. Li, T. Rindler-Daller, and P. R. Shapiro, ``Cosmological constraints on Bose-Einstein-condensed scalar field dark matter.'' Phys. Rev. D {\bf89} (2014) 083536. arXiv: 1310.6061
 \bibitem{Schive} H.-Y. Schive, T. Chiueh, and T. Broadhurst, ``Cosmic structure as the quantum interference of a coherent dark wave.`` Nature Phys. 10, 496-499 (2014). arXiv: 1406.6586
  \bibitem{Marsh} D. J. E. Marsh, ''Axion cosmology,'' Phys. Rept. 643, 1-79 (2016). arXiv: 1510.07633
     \bibitem{Witten} L. Hui, J. P. Ostriker, S. Tremaine, and E. Witten, ``Ultralight scalars as cosmological dark matter.'' Phys. Rev. D {\bf95} (2017) 043541. arXiv: 1610.08297
    \bibitem{JLee} J. Lee, ``Brief History of Ultra-light Scalar Dark Matter Models.'' EPJ Web of Conferences 168 (2018) 06005. arXiv: 1704.05057
 
 \bibitem{Sim1} H.-Y. Schive, T. Chiueh, and T. Broadhurst, “Cosmic Structure as the Quantum Interference of a Coherent
Dark Wave.” Nature Phys. 10 (2014) 496–499. arXiv: 1406.6586
 \bibitem{Sim2} H.-Y. Schive, M.-H. Liao, T.-P. Woo, S.-K. Wong, T. Chiueh, T. Broadhurst, and W. Y. P. Hwang, “Understanding the Core-Halo Relation of Quantum Wave Dark Matter from 3D Simulations.” Phys. Rev. Lett. {\bf113} no. 26, (2014) 261302. arXiv: 1407.7762
 \bibitem{Sim3} B. Schwabe, J. C. Niemeyer, and J. F. Engels, “Simulations of solitonic core mergers in ultralight
axion dark matter cosmologies.” Phys. Rev. D {\bf94} no. 4, (2016) 043513. arXiv: 1606.05151
 \bibitem{Sim4} J. Veltmaat and J. C. Niemeyer, “Cosmological particle-in-cell simulations with ultralight axion dark matter.” Phys. Rev. D {\bf94} no. 12, (2016) 123523. arXiv: 1608.00802
 \bibitem{Sim5} P. Mocz, M. Vogelsberger, V. H. Robles, J. Zavala, M. Boylan-Kolchin, A. Fialkov, and L. Hernquist, “Galaxy formation with BECDM: I. Turbulence and relaxation of idealized haloes.” Mon. Not. Roy. Astron. Soc. 471 no. 4, (2017) 4559–4570. arXiv: 1705.05845
 \bibitem{LymanAlpha} V. Irsic, M. Viel, M. G. Haehnelt, J. S. Bolton, and G. D. Becker, “First constraints on fuzzy dark matter from Lyman-$\a$ forest data and hydrodynamical simulations.” Phys. Rev. Lett. {\bf119} (2017) 031302. arXiv: 1703.04683
 \bibitem{BarBlum}  N. Bar, D. Blas, K. Blum, and S. Sibiryakov, "Galactic Rotation Curves vs. Ultra-Light Dark Matter: Implications of the Soliton -- Host Halo Relation." arXiv: 1805.00122
 \bibitem{Riotto} V. Desjacques, A. Kehagias, and A. Riotto, "The impact of ultra-light axion self-interactions on the large scale structure of the Universe." Phys. Rev. D {\bf97} (2018) 023529. arXiv: 1709.07946
 
  \bibitem{Backreaction} M. Rozner and V. Desjacques, "Backreaction of axion coherent oscillations." arXiv: 1804.10417
 \bibitem{GRB} J. Eby, P. Suranyi, and L.C.R. Wijewardhana, ''Expansion in Higher Harmonics of Boson Stars using a Generalized Ruffini-Bonazzola Approach, Part 1: Bound States.`` JCAP 1804 no. 04 (2018) 038. arXiv: 1712.04941
  
   \bibitem{Gupta} P. D. Gupta and E. Thareja, "Supermassive Black Holes from self-gravitating Bose-Einstein Condensates comprised of Ultra-light Bosonic Dark Matter." Classical and Quantum Gravity 34 (2017) 3. arXiv: 1512.08623
 
 \bibitem{GuthRelativistic} M. H. Namjoo, A. H. Guth, and D. I. Kaiser, "Relativistic Corrections to Nonrelativistic Effective Field Theories." arXiv: 1712.00445
 \bibitem{BraatenRelativistic} E. Braaten, A. Mohapatra, and H. Zhang, "Classical Nonrelativistic Effective Field Theories for a Real Scalar Field." arXiv: 1806.01898
 
 \bibitem{HeavySMB} O. Shemmer et al., "Near Infrared Spectroscopy of High Redshift Active Galactic Nuclei. I. A Metallicity-Accretion Rate Relationship." Astrophys. J. 614 (2014) 547-557. arXiv: astro-ph/0406559
 
 \bibitem{MWDM} G. Battaglia et al., "The radial velocity dispersion profile of the Galactic halo: Constraining the density profile of the dark halo of the Milky Way." MNRAS 364 (2005) 433-442; Erratum-ibid. 370 (2006) 1055. arXiv: astro-ph/0506102
 \bibitem{MWBH} S. Gillessen et al., "Monitoring stellar orbits around the Massive Black Hole in the Galactic Center."  Astrophysical Journal 692:1075-1109 (2009). arXiv: 0810.4674
 
  \bibitem{SchiveEvidence} I. De Martino, T. Broadhurst, S.-H. H. Tye, T. Chiueh, and H.-Y. Schive, "Dynamical Evidence of a Solitonic Core of $10^9M_\odot$ in the Milky Way." arXiv: 1807.08153
 
 \bibitem{AndromedaDM} P. R. Kafle et al., "The Need for Speed: Escape velocity and dynamical mass measurements of the Andromeda galaxy." MNRAS 475 (2018) 4043-4054. arXiv: 1801.03949
 \bibitem{AndromedaBH} R. Bender et al., "HST STIS spectroscopy of the triple nucleus of M31: two nested disks in Keplerian rotation around a Supermassive Black Hole." Astrophysical Journal 631 1 (2005). arXiv: astro-ph/0509839
  
 \bibitem{UCD1} M. Hilker, L. Infante, G. Vieira, M. Kissler-Patig, and T. Richtler, "The central region of the Fornax cluster. II. Spectroscopy and radial velocities of member and background galaxies." Astronomy and Astrophysics Supplement. 134 (1999) 75-86. arXiv: astro-ph/9807144
 \bibitem{UCD2} M. J. Drinkwater, J. B. Jones, M. D. Gregg, and S. Phillipps, "Compact Stellar Systems in the Fornax Cluster: Super-massive Star Clusters or Extremely Compact Dwarf Galaxies?" Publications of the Astronomical Society of Australia. 17 (2000) 227-233. arXiv: astro-ph/0002003
  \bibitem{UCDGC} S. Mieske, M. Hilker, and L. Infante, "Fornax compact object survey FCOS: On the nature of Ultra Compact Dwarf galaxies." Astron.Astrophys. 418 (2004) 445-458. arXiv: astro-ph/0401610
 \bibitem{UCDTidal1} M. J. Drinkwater et al., "A class of compact dwarf galaxies from disruptive processes in galaxy clusters." Nature 423 (2003) 519-521. arXiv: astro-ph/0306026
 \bibitem{UCDTidal2} K. Bekki, W. J. Couch, M. J. Drinkwater, and Y. Shioya, "Galaxy threshing and the origin of ultra-compact dwarf galaxies in the Fornax cluster." Mon. Not. Roy. Astron. Soc. 344 (2003) 399. arXiv: astro-ph/0308243
 
 \bibitem{UCDformation} A. V. Afanasiev et al., "A 3.5-million Solar Masses Black Hole in the Centre of the Ultracompact Dwarf Galaxy Fornax UCD3." MNRAS 477 (2018) 4. arXiv: 1804.02938
 
 \bibitem{UCDDensest} M. A. Sandoval et al., "Hiding in plain sight: record-breaking compact stellar systems in the Sloan Digital Sky Survey." The Astrophysical Journal Letters, Volume 808, Article L32, 2015. arXiv: 1506.08828
 
  \bibitem{Calabrese} E. Calabrese and D. N. Spergel, "Ultra-Light Dark Matter in Ultra-Faint Dwarf Galaxies." MNRAS 460 (2016) 4. arXiv: 1603.07321
 \bibitem{Du} X. Du, B. Schwabe, J. C. Niemeyer, and D. B\"urger, "Tidal disruption of fuzzy dark matter subhalo cores." Phys. Rev. D {\bf97} (2018) 063507. arXiv: 1801.04864
 
 \bibitem{UCDSMB} C. P. Ahn et al., "Detection of Supermassive Black Holes in Two Virgo Ultracompact Dwarf Galaxies". Astrophysical Journal. 839 (2017) 72. arXiv: 1703.09221
 \bibitem{Gaia} The Gaia Collaboration, "Gaia Data Release 2. Summary of the contents and survey properties." Astronomy and Astrophysics 616 (2018) A1. arXiv: 1804.09365
 
  \bibitem{EbyThesis} J. Eby, "Phenomenology and Astrophysics of Gravitationally-Bound Condensates of Axion-Like Particles." Ph.D thesis, University of Cincinnati. (2017). http://lss.fnal.gov/archive/thesis/2000/fermilab-thesis-2017-19.pdf.
 
  \bibitem{Stoof} H. Stoof, "Macroscopic quantum tunneling of a bose condensate." J. Stat. Phys. 87, 1353 (1997)
  \bibitem{Arovas} J. A. Freire and D. P. Arovas, "Collapse of a Bose condensate with attractive interactions." Phys. Rev. A {\bf59} (1999) 1461. arXiv: cond-mat/9803280
 \bibitem{Marquardt} K. Marquardt, P. Wieland, R. Häfner, H. Cartarius, J. Main, and G. Wunner, "Macroscopic quantum tunneling of Bose-Einstein condensates with long-range interaction." Phys. Rev. A {\bf86} (2012) 063629. arXiv: 1212.1316
 \bibitem{Coleman} Coleman and Callen Jr., "Fate of the false vacuum. II. First quantum corrections." Phys. Rev. D {\bf16} (1977) 1762.
 \bibitem{Milnikov} Mil’nikov and Nakamura, "Practical implementation of the instanton theory. II. Decay of metastable state through tunneling." J. Chem. Phys 117 (2002) 10081.
\end{thebibliography}
\end{document}